\documentclass[aps,prb,twocolumn,superscriptaddress,showpacs]{revtex4-1}
\usepackage{amsmath,amssymb,amsfonts,amsthm}
\usepackage{graphicx}
\usepackage{bm}
\usepackage{bbm}
\usepackage{bbold}
\usepackage{mathrsfs}
\usepackage{color}
\usepackage{dcolumn}   % needed for some tables
\usepackage{epstopdf}
\usepackage[caption=false]{subfig}
\usepackage{xr}
\usepackage[colorlinks=true,linkcolor=blue,urlcolor=blue,citecolor=blue]{hyperref}
\begin{document}

 \newcommand{\breite}{1.0} %  for twocolumn

\newtheorem{prop}{Proposition}
\newtheorem{cor}{Corollary} 

\newcommand{\be}{\begin{equation}}
\newcommand{\ee}{\end{equation}}

\newcommand{\bea}{\begin{eqnarray}}
\newcommand{\eea}{\end{eqnarray}}
\newcommand{\lt}{<}
\newcommand{\gt}{>} 

\newcommand{\Reals}{\mathbb{R}}     % Reals
\newcommand{\Com}{\mathbb{C}}       % Complex #
\newcommand{\Nat}{\mathbb{N}}       % Natural #

\newcommand{\id}{\mathbboldsymbol{1}}    

\newcommand{\Real}{\mathop{\mathrm{Re}}}
\newcommand{\Imag}{\mathop{\mathrm{Im}}}

\def\O{\mbox{$\mathcal{O}$}}   % Order epsilon ... 
\def\F{\mathcal{F}}			% FourierTrafo
\def\sgn{\text{sgn}}

\newcommand{\deo}{\ensuremath{\Delta_0}}
\newcommand{\dea}{\ensuremath{\Delta}}
\newcommand{\ak}{\ensuremath{a_k}}
\newcommand{\ad}{\ensuremath{a^{\dagger}_{-k}}}
\newcommand{\sx}{\ensuremath{\sigma_x}}
\newcommand{\sz}{\ensuremath{\sigma_z}}
\newcommand{\spl}{\ensuremath{\sigma_{+}}}
\newcommand{\smi}{\ensuremath{\sigma_{-}}}
\newcommand{\alk}{\ensuremath{\alpha_{k}}}
\newcommand{\bk}{\ensuremath{\beta_{k}}}
\newcommand{\ok}{\ensuremath{\omega_{k}}}
\newcommand{\vd}{\ensuremath{V^{\dagger}_1}}
\newcommand{\vi}{\ensuremath{V_1}}
\newcommand{\vo}{\ensuremath{V_o}}
\newcommand{\zc}{\ensuremath{\frac{E_z}{E}}}
\newcommand{\xc}{\ensuremath{\frac{\Delta}{E}}}
\newcommand{\xd}{\ensuremath{X^{\dagger}}}
\newcommand{\aok}{\ensuremath{\frac{\alk}{\ok}}}
\newcommand{\tpw}{\ensuremath{e^{i \ok s }}}
\newcommand{\tpe}{\ensuremath{e^{2iE s }}}
\newcommand{\tmw}{\ensuremath{e^{-i \ok s }}}
\newcommand{\tme}{\ensuremath{e^{-2iE s }}}
\newcommand{\epls}{\ensuremath{e^{F(s)}}}
\newcommand{\emis}{\ensuremath{e^{-F(s)}}}
\newcommand{\epl}{\ensuremath{e^{F(0)}}}
\newcommand{\emi}{\ensuremath{e^{F(0)}}}

\newcommand{\lr}[1]{\left( #1 \right)}
\newcommand{\lrs}[1]{\left( #1 \right)^2}
\newcommand{\lrb}[1]{\left< #1\right>}
\newcommand{\nbt}{\ensuremath{\lr{ \lr{n_k + 1} \tmw + n_k \tpw  }}}

\newcommand{\om}{\ensuremath{\omega}}
\newcommand{\dw}{\ensuremath{\Delta_0}}
\newcommand{\wbp}{\ensuremath{\omega_0}}
\newcommand{\dv}{\ensuremath{\Delta_0}}
\newcommand{\vbp}{\ensuremath{\nu_0}}
\newcommand{\vplus}{\ensuremath{\nu_{+}}}
\newcommand{\vminus}{\ensuremath{\nu_{-}}}
\newcommand{\wplus}{\ensuremath{\omega_{+}}}
\newcommand{\wminus}{\ensuremath{\omega_{-}}}
\newcommand{\uv}[1]{\ensuremath{\mathbf{\hat{#1}}}} % for unit vector
\newcommand{\abs}[1]{\left| #1 \right|} % for absolute value
\newcommand{\norm}[1]{\left \lVert #1 \right \rVert} % for absolute value
\newcommand{\avg}[1]{\left< #1 \right>} % for average
\let\underdot=\d % rename builtin command \d{} to \underdot{}
\renewcommand{\d}[2]{\frac{d #1}{d #2}} % for derivatives
\newcommand{\dd}[2]{\frac{d^2 #1}{d #2^2}} % for double derivatives
\newcommand{\pd}[2]{\frac{\partial #1}{\partial #2}} 
% for partial derivatives
\newcommand{\pdd}[2]{\frac{\partial^2 #1}{\partial #2^2}} 
% for double partial derivatives
\newcommand{\pdc}[3]{\left( \frac{\partial #1}{\partial #2}
 \right)_{#3}} % for thermodynamic partial derivatives
\newcommand{\ket}[1]{\left| #1 \right>} % for Dirac bras
\newcommand{\bra}[1]{\left< #1 \right|} % for Dirac kets
\newcommand{\braket}[2]{\left< #1 \vphantom{#2} \right|
 \left. #2 \vphantom{#1} \right>} % for Dirac brackets
\newcommand{\matrixel}[3]{\left< #1 \vphantom{#2#3} \right|
 #2 \left| #3 \vphantom{#1#2} \right>} % for Dirac matrix elements
\newcommand{\grad}[1]{{\nabla} {#1}} % for gradient
\let\divsymb=\div % rename builtin command \div to \divsymb
\renewcommand{\div}[1]{{\nabla} \cdot \boldsymbol{#1}} % for divergence
\newcommand{\curl}[1]{{\nabla} \times \boldsymbol{#1}} % for curl
\newcommand{\laplace}[1]{\nabla^2 \boldsymbol{#1}}
\newcommand{\vs}[1]{\boldsymbol{#1}}
\let\baraccent=\= % rename builtin command \= to \baraccent
%%%%%%%%%%%%%%%%%%%%%%%%%%%%%%%%%%%%%%%%%%%%%
% End Definitions
%%%%%%%%%%%%%%%%%%%%%%%%%%%%%%%%%%%%%%%%%%%%%

%\newcommand{\IM}[1]{\texttt{\textcolor{Green} #1}}
%\newcommand{\KA}[1]{\texttt{\textcolor{RoyalBlue} #1}}
\def\red#1{{\textcolor{red}{#1}}}
%\newcommand{\IMcomment}[1]{\texttt{\color{} #1}}

%Title of paper
\title{Quantum Hall valley Ferromagnets as a platform for topologically protected quantum memory}

\author{Kartiek Agarwal}
\email{agarwal@physics.mcgill.ca}
\affiliation{Department of Physics, McGill University, Montr\'{e}al, Qu\'{e}bec H3A 2T8, Canada}

\date{\today}
\begin{abstract}
Materials hosting topologically protected non-Abelian zero modes offer the exciting possibility of storing and manipulating quantum information in a manner that is protected from decoherence at the hardware level. In this work, we study the possibility of realizing such excitations along line defects in certain fractional quantum Hall states in multi-valley systems. Such line defects have been recently observed experimentally between valley polarized Hall states on the surface of Bi(111), and excitations near these defects appear to be gapped (gapless) depending on the presence (absence) of interaction-induced gapping perturbations constrained by momentum selection rules, while the position of defects is determined by strain. In this work, we use these selection rules to show that a hybrid structure involving a superlattice imposed on such a multi-valley quantum Hall surface realizes non-Abelian anyons which can then be braided by modulating strain locally to move line defects. Specifically, we explore such defects in Abelian fractional quantum Hall states of the form $\nu = 2/m$ using a K-matrix approach, and identify relevant gapping perturbations. Charged modes on these line defects remain gapped, while charge netural valley pseudospin modes may be gapped with the aid of two (mutually orthogonal) superlattices which pin non-commuting fields. When these superlattices are alternated along the line defect, non-Abelian zero modes result at points where the gapping perturbation changes. Given that these pseudospin modes carry no net physical charge or spin, the setup eschews utilizing superconducting and magnetic elements to engineer gapping perturbations. We provide a scheme to braid these modes using strain modulation, and confirm that the resulting unitaries satisfy a representation of the braid group. 
\end{abstract}
\maketitle

\section{Introduction}

Non-Abelian zero modes serve as a corner stone in the realization of topological quantum computing~\cite{kitaev2003fault,nayak2008non}. When physically isolated, these excitations encode a macrosopic ground state degeneracy that grows exponentially in their number, and which is robust to local perturbations. Furthermore, non-trivial unitary operations on this degenerate subspace can be realized by exchanging these excitations in what are known as braiding operations that are also topologically robust. Since the discovery of such excitations in the $\nu = 5/2$ fractional quantum Hall (FQH) state~\cite{moore1991nonabelions}, there has been an immense push both theoretically~\cite{read1996quasiholes,kitaev2001unpaired,ivanov2001non,kitaev2006anyons,fu2008superconducting,fu2009josephson,sau2010generic,lutchyn2010majorana,qi2010chiral,yang2016majorana,ren2019topological} and experimentally~\cite{mourik2012signatures,nadj2014observation,albrecht2016exponential} towards finding platforms that realize such modes, given their potential for quantum computing. 

Perhaps the most promising platform for the experimental realization of non-Abelian anyons comes in the form of Majorana zero modes (MZMs) realized in semiconducting wires with strong spin-orbit coupling and proximitized superconductivity, in the presence of strong magnetic fields~\cite{lutchyn2018majorana}. Another notable setting is that of line defects in bilayer systems supporting Abelian FQH states where local depletion of the electronic density using appropriate electrical gates and clever stitching of the two layers by inter-layer tunneling can result in an effective higher genus surface for the electrons of the system~\cite{barkeshli2014synthetic}. This also leads to the presence of non-Abelian excitations dubbed genons, which in turn encode the degeneracy of the FQH ground state on such a surface~\cite{wen1990ground}. 

Although much experimental progress has been made in realizing MZMs experimentally, engineering truly isolated MZMs good enough for decoherence-free qubit operations remains elusive. At least part of the difficulty there arises from the simultaneous presence of superconducting and magnetic correlations, which are deleterious to one another~\cite{flensberg2021engineered}, and disorder~\cite{ahn2020goodbadugly}. In the putative bilayer genon qubit, a key challenge arises from the balancing act between desiring a short inter-layer distance to allow for strong tunneling between the layers, and a larger inter-layer distance to prevent stray electric fields arising from gates coupled to one layer from penetrating into the other layer. 

In this work, we take inspiration from the theoretical proposals of genon qubits and the related proposal by Lindner et al.~\cite{lindner2012fractionalizing} for realizing fractionalized Majorana zero modes in Abelian FQH systems (see also related proposals~\cite{clarke2013exotic,cheng2012superconducting,vaezi2013fractional}), along with recent experimental developments in the realization of quantum Hall states in systems with valley degeneracy~\cite{feldman2016observation,randeria2019interacting,hossain2021bloch}, to propose a setup where non-Abelian zero modes can in principle be realized without simultaneously demanding superconducting and magnetic elements, and where braiding may be achieved with local strain modulation instead of electric gates. 

In effect, our proposal utilizes the multi-valley degeneracy of such systems to serve as a proxy for multiple layers of FQH states in the proposal for genon qubits. Line defects in these systems arise between regions with spontaneous but different selective populations of valleys in realizing FQH states. QH states have been observed in many multi-valley systems~\cite{dunford1995fqhe,lai2004two,lu2012fractional,dean2011multicomponent,feldman2012unconventional}, and more recently, valley ferromagnetism~\cite{feldman2016observation,hossain2021bloch} and line defects have also been observed in the integer filling case~\cite{randeria2019interacting}, and investigated theoretically~\cite{agarwal2019topology}. In the absence of point-like disorder that can trivially scatter electrons between valleys, and interactions, these line defects serve as edges for the valley polarized FQH states and carry valley-filtered edge modes in correspondence with the bulk state on either side of the defect. These defects themselves can be spatially mobile, unless they are pinned, for instance by a small strain gradient, that weakly breaks the valley degeneracy and favors a certain (and different) polarization on either side of the defect. Thus, these defects may be positionally controlled with weak strain modulation. 

When interactions are considered, electrons residing in these valley-filtered edge modes can scatter among each other, possibly leading to gapping perturbations. However, for these perturbations to be relevant, they must respect momentum conservation in the two-dimensional Brillouin zone (2DBZ). The presence or absence of these gapping perturbations is highly dependent on the bulk filling factor, which ultimately determines the valleys from which electrons form the edge modes---the presence of both gapless and gapped excitations have been observed in STM data~\cite{randeria2019interacting} and is in accordance with these theoretical expectations~\cite{agarwal2019topology}. 

In this work, we focus on a four-valley model (realized naturally on the surface of many semiconducting materials) and explore line defects when the bulk is described by an FQH state with filling factor $\nu = 2/m$, with $m$ being an odd integer. Using just the symmetries inherent to the valley degenerate system, we argue that the line defects can be described by an effective two-component Luttinger theory comprising an electrically charged `charge' mode, and an electrically neutral, `valley peudospin' mode. The charged mode turns out to be naturally gapped due to the presence of a certain scattering process that always respects momentum conservation in the 2DBZ. In analogy with Ref.~\cite{lindner2012fractionalizing}, if the pseudospin mode is now gapped with two distinct gapping perturbations that pin non-commuting fields, and which alternate along the line defect, non-Abelian zero modes are realized at the interface between such gapped regions. 
%As we show, these alternate gapping perturbations can be generated by imprinting a superlattice potential that allows for certain additional inter-valley scattering processes to become relevant. 

One can view these gapping perturbations as sutures that allow the free travel of certain quasiparticles in the FQH state across the line defect~\cite{barkeshli2013theory}. Accordingly, if the remaining pseudospin mode can be gapped via perturbations that pin non-commuting fields and alternate along the line defect, it is equivalent to creating a higher genus surface on which the Abelian FQH system resides, and concomitantly espouses a degeneracy exponential in the genus number. In the single layer context, when the quasiparticle modes are charged, one requires superconductivity and magnetic elements to realize such gapping perturbations~\cite{lindner2012fractionalizing}. Instead, the valley pseudospin modes are charge neutral (and spin polarized). As we show, relevant gapping perturbations can be generated by the joint action of an interaction-induced scattering event that does not conserve momentum in the 2DBZ by itself, in concert with an additional superlattice structure that resides on the line defect and can provide for the momentum mismatch. A static, imprinted superlattice structure with the appropriate geometry (to be specified later) can then be used to realize non-Abelian zero modes, while braiding is achieved via dynamical control of local strain fields. 

It is important to note that this proposal requires valley charges to be well defined, which prohibits the presence of point-like impurities that can scatter electrons from one valley to another. Small amounts of such impurities can of course be tolerated owing to the topological robustness of the underlying FQH state. This should be feasible in many of the two-dimensional materials and surfaces that espouse the valley degenerate structure we rely on, such as silicon, where immense technological advancements have been made at the industrial level to consistently produce high-mobility devices. Many elements of our proposal have been experimentally investigated, including QH states in multi-valley systems~\cite{dunford1995fqhe,lai2004two,lu2012fractional,dean2011multicomponent,feldman2012unconventional,feldman2016observation,hossain2021bloch}, line defects in such systems~\cite{randeria2019interacting}, controlling properties of valleys using strain modulation with piezoelectrics~\cite{sun2009strain,nityasagar2019valleydrift}, and imprinting of superlattices on high mobility two-dimensional electron gases~\cite{lu2016high,wang2021quantum,hossain2021bloch}. By themselves, multivalley quantum systems can engender a wide variety of FQH phenomena alongside conventional symmetry breaking orders with novel implications (see Ref.~\cite{parameswaran2019quantum} and references therein), and have immense potential for building novel devices and quantum engineering. Our work may thus be viewed as an addition to the rich tapestry of possibilities of interesting quantum phenomena in these systems. 

This manuscript is organized as follows. In Sec.~\ref{sec:summary}, we provide an intuitive discussion of our setup and main results. In particular, we outline the multi-valley system considered, the construction of a prototype topological qubit in this system, the appearance of ground state degeneracy toggled by non-Abelian zero modes, and a scheme using strain modulation to braid these zero modes. We draw connections and highlight distinctions between the genon qubit proposal and the multi-valley system at hand where relevant. In the following sections, we discuss these aspects in more detail. In Sec.~\ref{sec:bulk}, we detail the multi-valley system under consideration along with the form of interactions, and the possible bulk FQH states they could engender. In Sec.~\ref{sec:edge}, we detail the edge theory using the K-matrix formalism, the most general interactions allowed as per the symmetries of the system, and show how a two component Luttinger theory describes the behavior of the low-energy excitations. We then discuss the possible quasiparticle and electron operators realized and the gapping perturbations that can arise, including in the presence of a superlattice. In Sec.~\ref{sec:nonabelian}, we describe in detail the realization of non-Abelian quasiparticles in gapped line defects and their braiding properties. We conclude in Sec.~\ref{sec:conclusions} by briefly summarizing our findings. 
\begin{center}
    \begin{figure*}
    \includegraphics[width=0.9\textwidth]{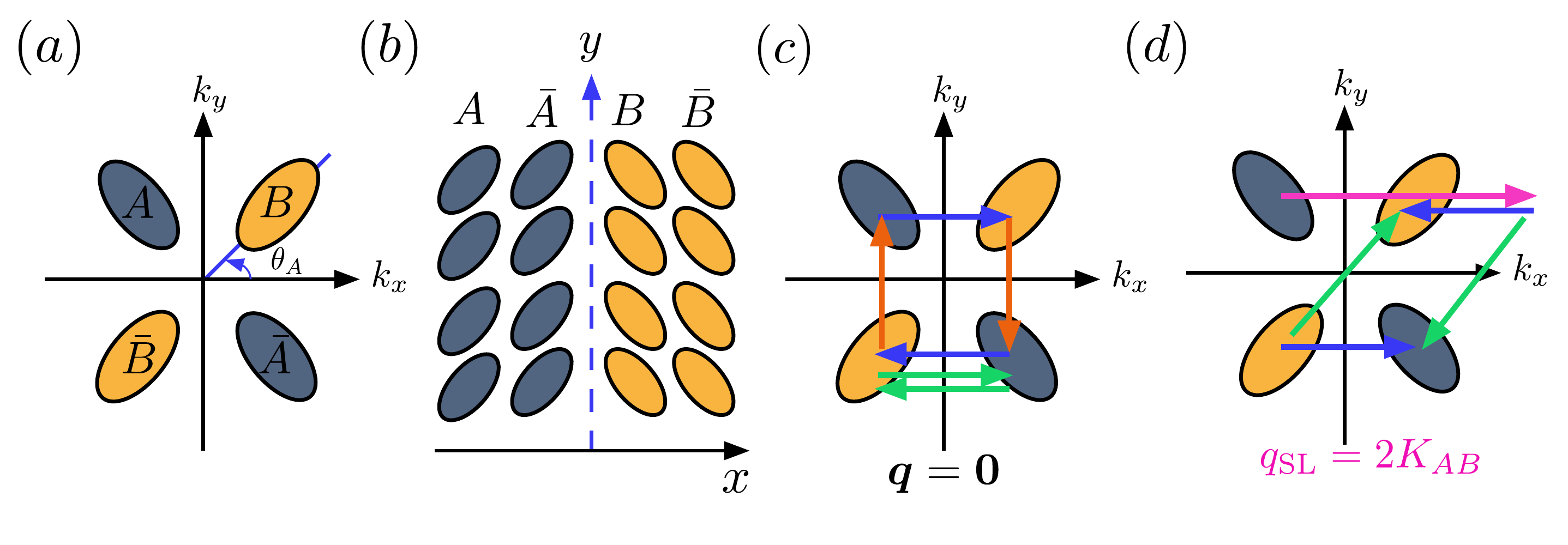}
    \caption{(a) The brilluoin zone of the 4-valley model in consideration with the valley pairs $A, \bar{A}$ (grey) and $B, \bar{B}$ (yellow) shown. b) A domain wall separating regions where FQH states with valleys $A, \bar{A}$ (partly) occupied for $x <0$ and $B, \bar{B}$ occupied for $x>0$. (c) Scattering processes on the line defect involving pairs of electrons are shown; colored arrow pairs indicate motion of a pair of electrons in the scattering process. (d) Some superlattice assisted scattering processes are shown; the momentum absorbed from the superlattice, $\vs{q}_{\text{SL}}$ is shown in magenta.}
    \label{fig:figbands}
    \end{figure*}
\end{center}
\section{Summary of proposal and main findings}
\label{sec:summary}

\subsection{4-valley model, bulk FQH state and line defects}

The simplest model that captures the physics we wish to elucidate is a 4-valley model [see Fig.~\ref{fig:figbands} (a) for an illustration of the 2DBZ in consideration] of spin-polarized electrons. We work in the regime where $\ell_B \gg a$ and $ K_{\alpha \beta} \ell_B \gg 1$, where $\ell_B = \sqrt{\hbar/eB}$ is the magnetic length, $a$ is the lattice constant, and $K_{\alpha \beta}$ is the wavevector separating the centers of valleys $\alpha$ and $\beta$. The valleys are related by discrete symmetries of the underlying crystal lattice and are degenerate. In this case, we assume that valleys $A, B$ and $\bar{A}, \bar{B}$ are related by mirror symmetry about an appropriate axis, while valleys $A, \bar{A}$ and $B, \bar{B}$ are related by a discrete $\pi$ rotation. Such a 2DBZ is, for instance, realized on a multitude of surfaces of various indirect band gap semiconductors and semimetals, such as silicon and bismuth, among others. The valleys usually have an anisotropic dispersion. This implies that a long-wavelength uniaxial strain modulation generically splits (unless it is perfectly aligned along the mirror axis between valleys $A$, $B$) the four-fold degeneracy into residual two-fold degeneracy of $A, \bar{A}$ and $B, \bar{B}$ valleys. In certain instances like Si(111) and Bi(111), another pair of valleys are realized; we will assume these to be split more strongly by uniform uniaxial strain from the remaining four valleys and thus need not be considered. 

We now envision preparing the system in a FQH state with filling factor $\nu = 2/m$, where $m$ is an odd integer. Let us first consider the integer case, $m = 1$. Suppose valleys $A, \bar{A}$ have an infinitesimally lower occupation energy (due to strain) in some region. In this case, an incompressible quantum Hall state arises with the zeroth Landau levels (LLs) of valleys $A, \bar{A}$ being occupied, and that of valleys $B, \bar{B}$, being empty---one can view this valley polarization as a consequence of Stoner's instability in a system where unlike the usual ferromagnetic case, there is no kinetic energy cost to selectively occupying one species of fermions as the magnetic field renders all orbitals in the zeroth LL equal in energy~\cite{abanin2010nematic,sondhi1993skyrmions}. [Occupation of just one valley ($A$ or $\bar{A}$) is energetically prohibitive at $\nu = 2$ due to the requirement of filling two LLs separated by the cylcotron gap.] The gap is then given by the Coulomb exchange gap, and is of the order of $e^2/\epsilon \ell_B$, where $\epsilon$ is the dielectric constant in the material. In the fractional case, $m \neq 1$, multiple candidate FQH states may arise. In this introductory section, we assume that the FQH state realized is a Halperin $(m,m,0)$ state~\cite{halperin1983theory}, corresponding to two copies of $\nu = 1/m$ Laughlin states~\cite{laughlin1983anomalous} in each valley $A, \bar{A}$. We will justify why such a state may be energetically preferred in the presence of a short enough screening length, by detailing more carefully the interactions in this system in Sec.~\ref{sec:bulk}; our analysis in proceeding sections will consider more general FQH states.

A line defect divides the system into two regions---one wherein valleys $A, \bar{A}$ are occupied, and another wherein valleys $B, \bar{B}$ are occupied; see Fig.~\ref{fig:figbands} (b). In the absence of point-like disorder, valley charges are separately conserved and can be thought of as $4$ independent $U(1)$ charges. The FQH state in the bulk then necessitates the presence of valley-filtered edge modes running in opposite directions along such a line defect---in particular, for the line defect at $x = 0$ shown in Fig.~\ref{fig:figbands}, edge modes carrying $1/m$ Laughlin quasiparticles sourced from the valley $A, \bar{A}$ run along the positive $y$ direction and $1/m$ quasiparticles sourced from valleys $B, \bar{B}$ run in the opposite direction. 

The properties of these edges are generically captured in a K-matrix description~\cite{wen1991topological,wen1992classification}. Then, provided there is no point-like disorder, the properties of the line defect are determined by interactions which scatter electrons between valleys. Among these, interactions that scatter electrons without changing their valley are the strongest as they do not involve any large momentum transfer (corresponding to an intervalley scattering wavevector). As we show, the $4$ edge modes are best described in terms of a two-component Luttinger theory with appropriate conjugate charge-current pairs given by 
\begin{align}
    \mathcal{Q_{\rho}} &= [\rho_A + \rho_{\bar{A}} + \rho_B + \rho_{\bar{B}}]/2 \nonumber \\
    \mathcal{J_{\rho}} &= [\rho_A + \rho_{\bar{A}} - (\rho_B + \rho_{\bar{B}})]/2 \nonumber \\
    \mathcal{Q_{\sigma}} &= [\rho_A - \rho_{\bar{A}} + \rho_B - \rho_{\bar{B}}]/2 \nonumber \\
    \mathcal{J_{\sigma}} &= [\rho_A - \rho_{\bar{A}} - (\rho_B - \rho_{\bar{B}})]/2
    \label{eq:charges}
\end{align}
Here, $\rho_\alpha$ is the density of electrons in valley $\alpha$, and $\mathcal{Q_\rho}, \mathcal{J_\rho}$ are identified as the conjugate charge-current pair corresponding to the charge mode, and $\mathcal{Q_\sigma}, \mathcal{J_\sigma}$ as the conjugate pair corresponding to the valley pseudospin mode. The latter clearly does not carry net electrical charge, and when all electrons are spin polarized, does not, effectively, carry physical spin. 

We now discuss the possible gapping perturbations on the line defect. These involve electrons tunneling between valleys, and are illustrated in Fig.~\ref{fig:figbands} (c). As an example, the green arrows correspond to an intervalley scattering between valleys $\bar{A}, \bar{B}$. Such an interaction is allowed as the net momentum transfer in the 2DBZ is zero, but it does not lead to a gapping perturbation since it preserves valley occupation numbers and can be thus subsumed into the effective Luttinger liquid description with renormalized parameters. The only gapping perturbation is described by the pair of blue and orange arrows. Both these processes correspond to a net transfer for electrons from up moving $A, \bar{A}$ modes into down moving $B, \bar{B}$ modes. This term is captured by a gapping sine-Gordon term of the form $\cos (2 m \phi_\rho)$, where $\mathcal{Q_\rho} = \nabla \phi_\rho / 2\pi$, in the Luttinger theory. This perturbation is always relevant when the Luttinger parameter $K_\rho$ is sufficiently small; as we will show this is generically the case in the limit of weak strain gradient that ultimately pins the line defect. The resultant gap has been observed in STM measurements on Bi(111) for the integer $m=1$ case and stands in contrast to an edge mode realized in the same physical location when the filling factor is $\nu = 1$; due to the absence of electrons in all $4$ valleys, the above interaction becomes irrelevant and no excitation gap is observed. 

We note that the valley pseudospin mode does not have natural gapping perturbations as these correspond to net momentum transfer in the 2DBZ and are exponentially suppressed as $e^{- 2K_{AB} \ell_B}$. (Landau level projection in principle allows for a process in the 2DBZ that does not conserve net momentum but it is exponentially weak in the momentum mismatch.) We now envision regions where a superlattice is imposed on top of the line defect, described by a wave-vector $\vs{q}_{\text{SL}} = 2 \vs{K}_{AB}$ and provides this remaining momentum in a perturbative manner; see Fig.~\ref{fig:figbands} (d). A gapping perturbation of the form $\cos (2 m \phi_\sigma)$ can now appear, which sends $\mathcal{J_\sigma} \rightarrow \mathcal{J_\sigma} \pm 2$ while keep other charges unchanged. Here, $\mathcal{J_\sigma} = \nabla \theta_\sigma / 2 \pi$. When this perturbation is sufficiently strong, it can open up a gap, and identifies charge $\mathcal{J_\sigma}$ modulo $2$.  Similarly, when a superlattice with wavevector $\vs{q}_{\text{SL}} = 2\vs{K}_{A\bar{B}}$ is introduced, it leads to a perturbation $\cos (2 m \theta_\sigma)$, which identifies $\mathcal{Q_\sigma}$ modulo $2$. We assume these two perturbations are applied in a spatially alternating way along the line defect, and next examine the consequences.

\subsection{Presence of non-Abelian zero modes}

The presence of non-Abelian zero modes in this system can be surmised by writing down a complete set of non-commuting Wilson loops~\cite{barkeshliqi1,barkeshliqi2,tong2016quantum}. These loops commute with the FQH Hamiltonian trivially. The minimum dimension of the irreducible representation corresponding to the algebra they satisfy then determines the minimum possible degeneracy of the ground state. The presence of non-commuting Wilson loops is, in fact, inherently tied to the genus of the surface on which the Abelian FQH state resides. The presence of zero modes can also be understood by examining appropriate Wilson lines. 

We first discuss Wilson loops in the original genon qubit platform. This is a bilayer system where a line defect can be thought to have cut the bilayer system in half. We assume that $\nu = 1/m$ FQH states are realized on both the layers. There are two relevant gapping perturbations in this case---i) electron tunneling between the left and right halves of the bilayer system, while remaining in the same (top or bottom) layer, and ii) electron tunneling between left half of the bottom layer to the right half of the top layer, and vice versa. This situation is formally equivalent to our multivalley setup with the following difference---the gapping perturbations are single particle tunneling events, as opposed to interactions that involve two particle processes in the multivalley system we study. 

\begin{center}
    \begin{figure}
    \includegraphics[width=3.3 in]{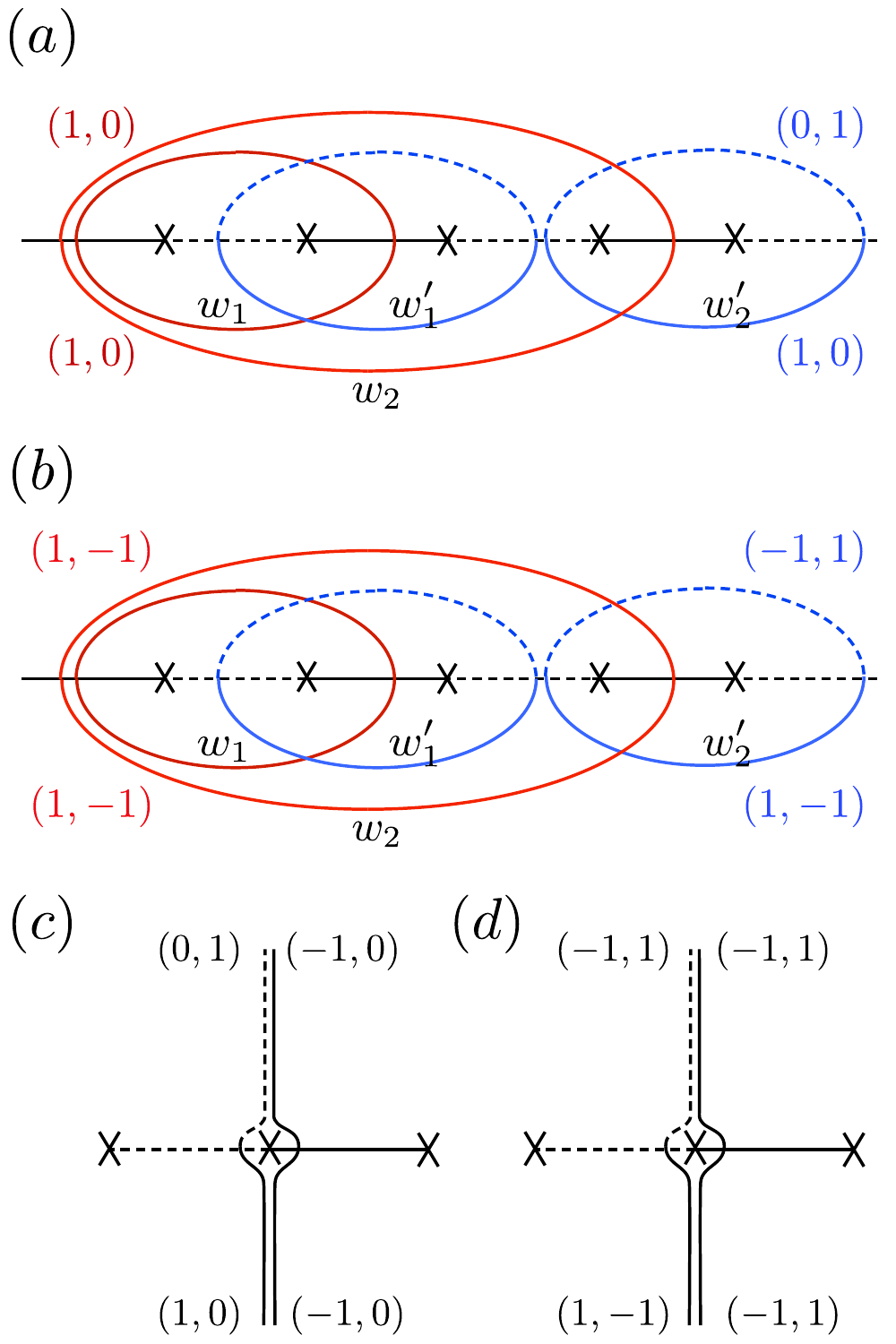}
    \caption{(a) and (b) show non-commuting pairs of Wilson loop operators in the bilayer genon qubit case and single layer multivalley case. The line defect is shown as a black solid and dashed line indicating the alternating form of gapping perturbation along the defect. See main text for an interpretation of the numbers associated. (c) and (d) show a pair of quasiparticles moving around an interface marked by an 'x' on the line defect for (a) and (b) respectively.}
    \label{fig:wilsonloop}
    \end{figure}
\end{center}
The Wilson loop is defined as the exponent of the line integral of the emergent Chern-Simons gauge field, $w_{\mathcal{C}} (l_I) = e^{i \int_{\mathcal{C}} l_I a_{I,\mu} d l^\mu}$, where $l_I$ is a vector of integers representing the (quantized) charge of the quasiparticle being moved along the directed curve $\mathcal{C}$. In Fig.~\ref{fig:wilsonloop}, we highlight the relevant pairs of non-commuting Wilson loops, in red and blue. Here, the top and bottom halves of both parts (a) and (b) of the figure represent regions on either side of the line defect, and the black solid and dashed lines represent the two alternating gapping perturbations. In particular, in part (a), the solid line indicates the tunneling perturbation suturing the top and bottom layers (separately), and the dashed line indicates the perturbation suturing the top layer to the bottom layer and vice-versa. The charge vector $l_I$ corresponding to these loops is also shown---$(1,0)$ and $(0,1)$ reflect, in their respective regions whether we are considering a Laughlin $1/m$ quasiparticle in the top layer [(1,0)], or the bottom layer [(0,1)]. The dashed blue line reflects the fact that a quasiparticle from the top layer coherently tunnels into the bottom layer when it crosses the line defect in a region when the interlayer tunneling is prevalent. These Wilson loops trivially commute with the Hamiltonian of the system. Importantly, in this case, they can be arranged into pairs of non-commuting operators. Specifically, one can easily show that $w_n w'_{n'}= \delta_{n,n'} e^{2\pi i/m} w'_{n'} w_n$. Thus, if there are $2N$ gapping regions, we have $N-1$ independent pairs of non-commuting Wilson loops (assuming the line defect winds around to form a circle). Since for each pair, the minimal irreducible representation is of dimension $m$, the topologically protected ground state degeneracy is $m^{N-1}$. 

One can also understand the presence of non-Abelian zero modes at the interface between these alternate gapping perturbations as follows; these are marked by an '\text{x}' in Fig.~\ref{fig:wilsonloop}. Imagine gently separating a quasiparticle $(1,0)$ and a quasihole $(-1,0)$ pair below the defect [see Fig.~\ref{fig:wilsonloop} (c) for illustration]. If one tunnels the quasiparticle across the line defect to the immediate left of an interface, it remains a $(1,0)$ quasiparticle. On the other hand, if we tunnel the $(-1,0)$ quasihole to the immediate right of the same interface, it tunnels across as a $(0,-1)$ quasihole. Thus, one can view the interface as a zero cost source of $(1,-1)$ quasiparticles. This intuitively explains the presence of non-Abelian zero modes at the interface. 

The multivalley system is different due to the fact that the suturing of the left and right halves of the system is performed by interaction processes instead of single particle scattering events. In this case, one gapping perturbation, in the case where the superlattice provides wavevector $\vs{q}_{\text{SL}} = 2 \vs{K}_{AB}$, converts a quasiparticle of valley A, and quasihole of valley $\bar{A}$, represented as $(1,-1)$, to a quasiparticle-quasihole pair in valleys $B, \bar{B}$ also represented as $(1,-1)$. On the other hand, the gapping perturbation corresponding to the superlattice wavevector $\vs{q}_{\text{SL}} = 2 \vs{K}_{A \bar{B}}$ converts a $(1,-1)$ quasiparticle to a $(-1,1)$ quasiparticle on the other side. [See Fig.~\ref{fig:wilsonloop} (d) for illustration.] One can again examine these Wilson loops to arrive at the following algebra: $w_n w'_{n'}= \delta_{n,n'} e^{8\pi i/m} w'_{n'} w_n$, while other pairs of loops commute with one another. This algebra again has a minimal irreducible representation of dimension $m$, and the ground state degeneracy is $m^{N-1}$ for $2N$ interfaces. The quantum dimension of the non-Abelian anyons realized is $\sqrt{m}$ as well in this case, akin to the case of the genon qubit. [In this analysis we ignored the fact that charge channel is also gapped, but using this fact one can tack on the coherent tunneling of the appropriate linear combination of charges in the charge channel along with the Wilson loops considered in Fig.~\ref{fig:wilsonloop} (b) to produce the same algebra as shown in Fig.~\ref{fig:wilsonloop} (a); thus topologically, the two cases are the same.]

We note in passing that a very similar setup is used in Ref.~\cite{lindner2012fractionalizing,clarke2013exotic} in a single layer system, with one side of the line defect corresponding to spin up electrons residing in a Laughlin $1/m$ state, and the other side corresponding to spin down electrons in the same $1/m$ state. The alternate gapping perturbations here arise from superconducting and magnetic perturbations. The non-Abelian zero modes realized in this system have quantum dimension $\sqrt{2m}$ due to the additional particle-hole symmetry imposed by the superconductor. One can view the Hilbert space of the ground state as being a tensor product of that engendered by non-Abelian anyons of quantum dimension $\sqrt{m}$ along with Majorana zero modes of quantum dimension $\sqrt{2}$. 

\subsection{Schematic model of topological qubit, zero mode operators, and braiding scheme}

\begin{center}
    \begin{figure}
    \includegraphics[width=2.7in]{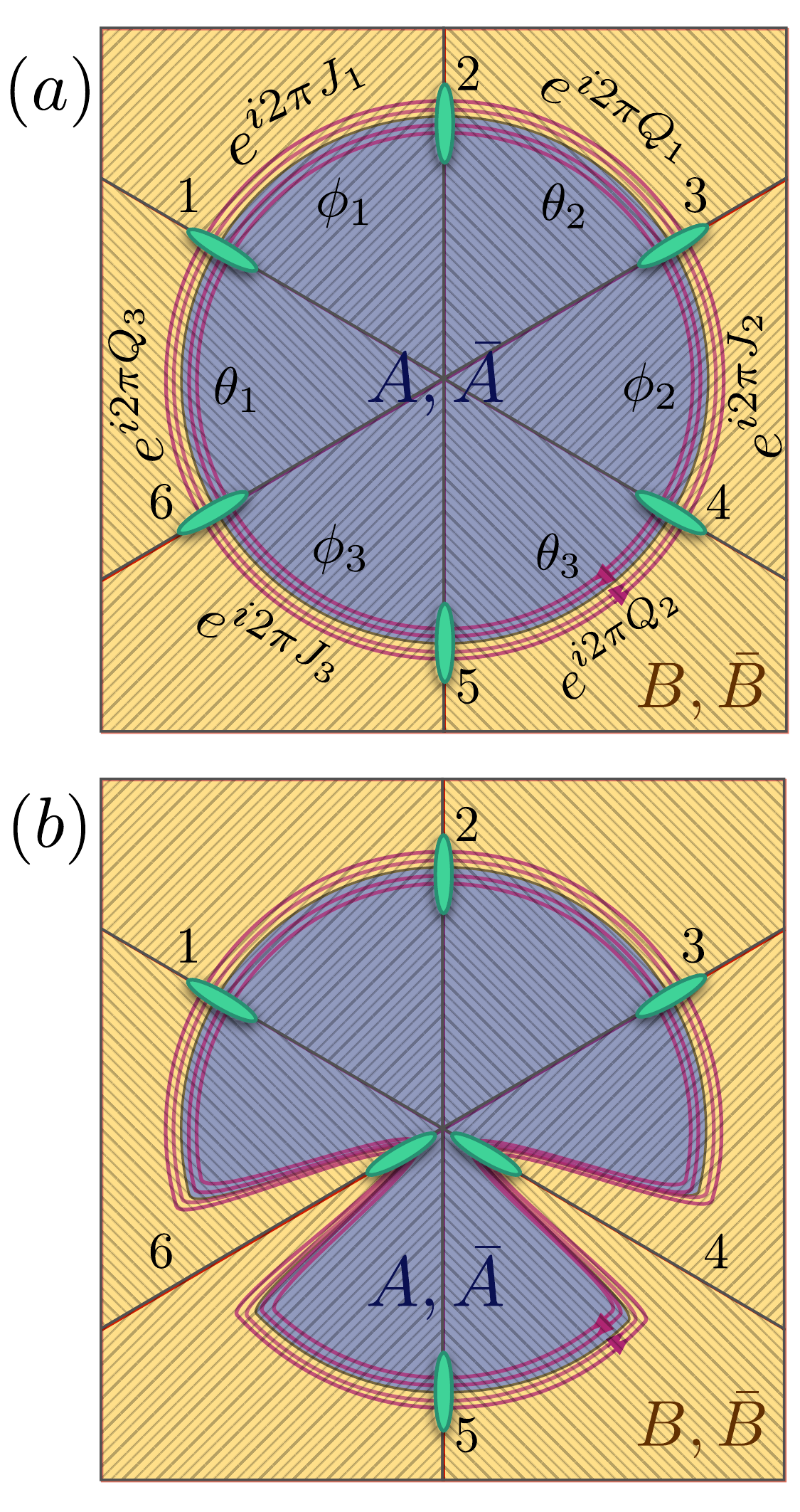}
    \caption{(a) A prototype topological qubit on the quantum Hall multivalley platform. The grey region indicates occupation of valleys $A, \bar{A}$ in a bulk FQH state, while the yellow region corresponds to the occupation of valleys $B, \bar{B}$. The line defect (shown in red) realizes a two-component Luttinger liquid. Hatching patterns divide this system into $6$ equal slices; they alternate in space and reflect two mutually orthogonal superlattice perturbations. These perturbations pin the fields $\phi_\sigma, \theta_\sigma$ where appropriate (with values denoted by $\phi_1, \theta_1, \dots$). Segments of the line defect can be characterized by the expectation values of the charges $Q_i, J_i$ defined modulo $1$. At the interfaces, non-abelian zero modes are realized; these are shown in green. (b) Zero modes can be brought into contact by `pinching' the region supporting $A, \bar{A}$ valleys along a radial line at the relevant interfaces. This could be achieved by strain modulation which can gently bias the system energetically towards supporting valley ferromagnetism in the chosen valley pair.}
    \label{fig:prototype}
    \end{figure}
\end{center}

We now turn to providing a schematic of a topological qubit realized in this system, as shown in Fig.~\ref{fig:prototype} (a). In this particular geometry, we imagine a superlattice structure with periodicity governed by wave-vectors $\vs{q}_{\text{SL}} = 2 \vs{K}_{AB}, 2 \vs{K}_{A\bar{B}}$ alternating in space and dividing the two-dimensional system into $6$ equal slices (shown as orthogonal hatching patterns). Small piezoelectric couplers control the local strain throughout this two-dimensional system, and in particular, support a FQH state with valleys $A, \bar{A}$ occupied inside a circular region in the middle while valleys $B, \bar{B}$ are occupied on the outside of this circular region. The periphery of the circular region then supports the two-component Luttinger liquid discussed above, with a uniformly gapped charge mode, and a valley pseudospin mode whose gapping perturbation varies along the line cut. This variation of the gaping perturbation induces non-Abelian zero modes (shown as green ellipsoids) at the interfaces where the gapping perturbation (the superlattice orientation) changes. 

The gapping perturbations $\cos (2m \theta_\sigma), \cos (2m \phi_\sigma)$ pin the fields $\theta_\sigma, \phi_\sigma$ in the appropriate regions, and identify the charges $Q_\sigma, J_\sigma$ modulo $2$, respectively. In what follows, we will denote the pinned values of these fields by $\phi_1, \phi_2, \dots, \theta_1, \theta_2, \dots$, as in Fig.~\ref{fig:prototype}, and forgo the subscript $\sigma$ for brevity. These fields in particular define the charges $2 \pi Q_i = \phi_{i+1} - \phi_i, 2 \pi J_i = \theta_{i+1} - \theta_{i}$ trapped in the segment between the two regions where these fields occupy definite minima of the cosine potential. Note that the minima are separated by a translation of $\phi, \theta$ by $\pi/m$, which corresponds to a change in the charges $Q_i, J_i$ by $1/2m$. Physically, this translation corresponds to the tunneling in of a $1/m$ Laughlin quasiparticle in one of the valleys [see the definition of these charges in Eq.~(\ref{eq:charges})] into this segment. However, it is important to note that such a change also affects the gapped charge mode of the two-component Luttinger system, as $Q_\rho \rightarrow Q_\rho \pm 1/2m$. To remain in the ground state, we must only consider changes that leave the charges $Q_\rho$ unchanged and $J_\rho$ unchanged moduolo $2$. This corresponds to tunneling in of a quasiparticle-quasihole pair with a net zero electrical charge, and translates into a minimal change in $Q_\sigma, J_\sigma$ by at least $1/m$ instead of $1/2m$.
%, and concomitantly, a minimal translation in $\phi_i, \theta_i$ by $2\pi/m$. 
Finally, we note that fields $\phi_i$ and $\theta_j$ cannot be specified simultaneously as they do not commute with one another. Accounting for conservation of total $\mathcal{Q_\sigma}, \mathcal{J_\sigma}$ on the line defect, this leaves $N-1$ independent $\phi_i$. The ground state degeneracy is given by counting the physically distinct minima $\phi_i$ can occupy. 

It is tempting to come to the conclusion that there are $2m$ distinct physical minima for each $\phi_i$---translating $\phi_i \rightarrow \phi_i + 2\pi/m$ changes appropriate $Q_i \rightarrow Q_i \pm 1/m$; since these $Q_i$ are identified modulo $2$, we may expect $2m$ unique ground states labeled uniquely by the expectation value $e^{i \pi Q_i}$. This would lead to a ground state degeneracy of $(2m)^{N-1}$, contrary to the result of $m^{N-1}$ found above using the algebra of Wilson loops. One can resolve this issue as follows. Note that a change of a pair of $Q_i \rightarrow Q_i + 1, Q_j \rightarrow Q_j - 1 \equiv Q_j + 1$ is a trivial operation. The edge theory is ultimately coupled to the bulk and extra charges on the edge are sourced from the bulk. Changing $Q_i \rightarrow Q_i + 1$ for instance involves sourcing an electron in valley $A$ from the bulk and pushing an electron in valley $\bar{A}$ into the bulk [see Eq.~(\ref{eq:charges})]; on the other hand, the change $Q_j \rightarrow Q_j - 1$ entails the exact opposite. Such an operation leaves the bulk unchanged (particularly because these are not fractional charges which require exciting the bulk). One can see this mathematically as well in terms of the commutation relations of edge operators. For instance, the operator $e^{i 2\pi m Q_i}$ sends $J_i \rightarrow J_i - 1, J_{i+1} \rightarrow J_{i+1} + 1$ but it is clearly an identity operator \emph{on the ground state subspace} where the charge $Q_i$ is quantized in units of $1/m$ (as noted above). One can run the argument in reverse showing that a change $Q_i \rightarrow Q_i + 1, Q_j \rightarrow Q_j - 1$ is indeed a trivial operation. Now, if each of $N$ $Q_i$ can take $2m$ values, the ground state degeneracy would be $(2m)^N$. However, there are $N-1$ pair constraints identified above that reduce this degeneracy to $(2m)^N/2^{N-1} = 2 m^N$. Finally, since the total charge is conserved modulo $2$ (and represents a condition independent of pair constraints which do not affect the total charge), working in a fixed charge sector yields the net degeneracy of $2 m^N / (2m) = m^{N-1}$, in agreement with the Wilson loop arguments. In general, we can choose a set of $N-1$ independent charges $Q_1, \dots, Q_{N-1}$ that are defined modulo $1$ using the above symmetry operations to represent the ground state subspace; alternatively, we can label each segment by the expectation value of the operator $e^{i 2\pi Q_i}$, with $Q_i = q_i/m$ and $q_i = 1, \dots, m-1$.

The non-Abelian zero modes realized in this system are effectively local pairs of quasiparticle operators projected onto the ground state manifold. These operators toggle between the various ground states. For instance, if we want to change $Q_i \rightarrow Q_i + 1/m$, we can apply an operator $\chi_A (x) \sim \psi^\dagger_A (x) \psi_{\bar{A}} (x) \sim e^{i \phi_\sigma (x) + i \theta_\sigma (x)}$, for some $x$ inside the region of interest. Similarly, we could apply an operator $\chi_B (x) \sim \psi^\dagger_B (x) \psi_{\bar{B}} (x) \sim e^{i \phi_\sigma (x) - i \theta_\sigma (x)}$. 
%Tunneling of $1/m$ quasiparticles from valley $A, \bar{A}$ to valley $B, \bar{B}$ across the line defect is disallowed as it costs energy and we thus need not consider other charge netural operators. 
Of course, since the line segment is gapped, these operators cost energy, and matrix elements of these operators projected onto the ground state subspace only have weight at the interface between segments gapped by alternating perturbations; this can be surmised from the fact that correlators of at least one of $\theta_\sigma$ and $\phi_\sigma$ fields will decay on either side of the interface (while the conjugate field is pinned and has longer range correlations). These form the zero modes of our system and can be shown to satisfy non-trivial commutation relations. 

%(It is worth noting that there are in fact $4$ other flavors of zero modes realized in this system, corresponding to projections of quasiparticle-quasihole operators from all possible valley pairs $\kappa, \kappa'$ with $\kappa \neq \kappa'$. The remaining $4$ operators additionally toggle between inequivalent but degenerate minima in the charge channel. They do not modify our main findings and are discussed in more detail in Appendix~\ref{sec:flavors}.)  

The zero modes cannot acquire finite expectation values by themsleves as this involves finite and large momentum transfer in the 2DBZ that is not supplied by the imposed superlattice structure. However, when brought close to one another, by way of bringing the interfaces closer, they can couple to other zero modes of the same flavor, and terms such as $\sim \chi^\dagger_\uparrow (x) \chi_\uparrow (x')$ can arise. In the prototype shown in Fig.~\ref{fig:prototype}, we envision bringing these zero modes closer together by pinching the region where the FQH state in valleys $A, \bar{A}$ is realized, to a smaller region. This can be done, for instance, by dynamically modulating the local strain field to move the line defect inwards towards the center. One can imagine bringing any pair of zero modes stuck at the interface closer in this manner.  

% Fig.~\ref{fig:prototype} (b) that in the braiding process we envision, $A,\bar{A}$ edge modes are brought closer to one another than $B, \bar{B}$ edge modes at all times. The latter can thus be exponentially weaker and neglected in this process. 
%
Following Ref.~\cite{lindner2012fractionalizing}, we can define a scheme using these processes to braid the zero modes. The scheme involves access to a resource vacuum that can always be used to furnish a pair of zero modes. For instance, in Fig.~\ref{fig:prototype}, we can imagine that the zero modes at interfaces $2,3$ are usually held close together by pinching; in this way, they can serve as a vacuum resource. Then, unpinching the modes creates a pair of zero modes from this vacuum with a definite charge $q_1$. To braid the zero modes, say $4,5$, we `copy' mode $4$ on to mode $2$ by pinching modes $3,4$ together. We then copy mode $5$ on to mode $4$ by pinching modes $3,5$ together. And finally, by pinching modes $2,3$ together, mode $2$ is copied on to mode $5$, thus finishing the exchange of modes $4,5$. We study such a braiding scheme and the corresponding unitary realized for our model. 

If we denote the unitary for braiding modes at the interfaces $2i,2i+1$, as $U_{2i,2i+1}$, we find 
\begin{align}
\mathcal{U}^{(k)}_{2i,2i+1} &= e^{-i \pi m \left( \mathcal{Q}_{i} - \frac{k}{2m}\right)^2} \nonumber \\
\mathcal{U}^{(k')}_{2i-1,2i} &= e^{-i \pi m \left( \mathcal{J}_{i} - \frac{k'}{2m}\right)^2}, 
\label{eq:braidU}
\end{align}
where $k, k'$ are some odd integers that depend on the microscopic implementation of tunneling between the zero modes. Thus, the braiding operations realize state dependent transformations in the ground state subspace. We also confirm that 

\begin{align}
    \left[ U^{(k_i)}_{i,i+1},U^{(k_j)}_{j,j+1} \right] &= 0 \; \; \text{for} \; \; |i-j| > 1,\nonumber \\
    U^{(k_1)}_{i,i+1} U^{(k_2)}_{i+1,i+2} U^{(k_1)}_{i,i+1} &= U^{(k_2)}_{i+1,i+2} U^{(k_1)}_{i,i+1} U^{(k_2)}_{i+1,i+2},
\end{align}

or the fact that the braiding transformations commute if they do not involve the same zero mode, and satisfy the Yang-Baxter equation, thus forming a representation of the braid group. 

We note that the above discussion was based on the assumption that the FQH state realized is of the Halperin $(m,m,0)$ kind. This is not guaranteed and needs detailed numerical simulations beyond the scope of this work. However, we provide some arguments in Sec.~\ref{sec:bulk} how such a state may be stabilized in the presence of sufficiently strong screening. We also consider more general FQH states of the form $(m',m',n)$ with filling fraction $\nu = 2/m$. The quantum dimension of the non-Abelian anyons realized in this case is found to be $\sqrt{\abs{m'-n}}$, and the braiding unitaries realized, as shown in Eq.~(\ref{eq:braidU}) can be obtained by changing $m \rightarrow \abs{m'-n}$. 

\section{Bulk model}
\label{sec:bulk}

We first describe the microscopic model governing the multi-valley system. The purpose of this section is to discuss in more detail the kinds of interactions that arise in this system when projected on to the lowest Landau level, and to motivate the possible fractional quantum Hall states that such a system may harbor. 

As noted above, we study a $4$-valley model 
(Fig.~\ref{fig:figbands}) of spin-polarized electrons described in a continuum effective mass approximation (valid when $\lambda_F, \ell_B\gg a $, where $\lambda_F$ is the Fermi wavelength, $a$ is of the order of the lattice spacing, and $\ell_B = \sqrt{\hbar/eB}$ is the magnetic length). In particular, it is also possible that the point-group symmetry of a material houses $6$ degenerate valleys; this occurs for instance on the $(111)$ surfaces of both Silicon and Bismuth. In this case, the anisotropy of the valleys can be used to lift the degeneracy using strain and reduce it to the $4$-fold degeneracy we study here (as was the case in the experiments of Ref.~\cite{randeria2019interacting}). In general the valleys are related by point group symmetries of the underlying lattice. Here we assume that valley pairs $\kappa, \bar{\kappa}$ are related by $\pi$-rotation, while valleys $A,B$ and $\bar{A}, \bar{B}$ are related by reflection about a high symmetry axis.  

The single-particle Hamiltonian for valley $\alpha \in \{A, B, \bar{A}, \bar{B} \}$ can be approximated as 
\begin{align}
H_\alpha = \frac{(p_\parallel - K + e A_\parallel / c)^2}{2 m_\parallel} + \frac{(p_\perp + e A_\perp / c)^2}{2 m_\perp},
\end{align}
 where $v_\parallel = v_x \cos \theta_\alpha + v_y \sin \theta_\alpha$, $v_\perp = v_y \cos \theta_\alpha - v_x \sin \theta_\alpha$ for any vector $\vs{v}$, and 
 $\theta_\alpha$ is the angle to the valley center from the $k_x$ axis. The valleys are centered at
 %\begin{align}
 $\vs{K}_\alpha = K (\cos \theta_\alpha, \sin \theta_\alpha)$, %\nonumber \\
 where we define $\vs{K}_{\alpha \beta} = \vs{K}_\alpha - \vs{K}_\beta$.
 %\end{align} %The valley dispersion need not be 
 We assume that deviations from ellipticity (e.g., from the teardrop shape of Bi(111) valleys) denoted $\delta H_\alpha$,  are smaller than the mass anistropy $\lambda^2 = m_\parallel/m_\perp$. Working in Landau gauge $\vs{A} = (0, Bx)$, and introducing a guiding center coordinate $X$ related  to the momentum via  $X = \ell^2_B p_y$, yields single-particle wavefunctions $\phi_{\alpha, X} (\vs{r})$  in valley $\alpha$ 
 \begin{equation}\label{eq:LLLorb}
\phi_{\alpha, X} (x,y) = \frac{e^{i X y + i \vs{K}_\alpha \cdot \vs{r} }}{\sqrt{L_y}} \left( \frac{z'_\alpha}{\pi} \right)^{1/4} e^{-\frac{z_\alpha (x + X)^2}{2}}, 
\end{equation}
where $L_y$ is the length of the QH sample in the $y$-direction, $\lambda^2 = m_\parallel/m_\perp$ is the mass anisotropy, $z_\alpha = \frac{\lambda}{\lambda^2 \sin^2 \theta_\alpha + \cos^2 \theta_\alpha} + i \frac{\sin 2 \theta_\alpha ( 1- \lambda^2)}{2 (\lambda^2 \sin^2\theta_\alpha + \cos^2 \theta_\alpha)}$, and $z'_\alpha = \text{Re} \left[ z_\alpha \right]$.

%(See Supplemental Material, Ref.~\cite{SupMat}).
 
%  For  $\delta H_\alpha=0$ the  LLL eigenfunctions in valley $\alpha$ is ($\ell_B=\hbar=1$ henceforth)
%\begin{equation}\label{eq:LLLorb}
%\phi_{\alpha, X} (x,y) = \frac{e^{i X y + i \vs{K}_\alpha \cdot \vs{r} }}{\sqrt{L_y}} \left( \frac{z'_\alpha}{\pi} \right)^{1/4} e^{-\frac{z_\alpha (x + X)^2}{2}}, 
%\end{equation}
%where $L_y$ is the length of the QH sample in the $y$-direction, $\lambda^2 = m_\parallel/m_\perp$ is the mass anisotropy, 
%$z_\alpha = \frac{\lambda}{\lambda^2 \sin^2 \theta_\alpha + \cos^2 \theta_\alpha} + i \frac{\sin 2 \theta_\alpha ( 1- \lambda^2)}{2 (\lambda^2 \sin^2\theta_\alpha + \cos^2 \theta_\alpha)}$, and $z'_\alpha = \text{Re} \left[ z_\alpha \right]$. 

Each non-interacting LL has an exact four-fold valley degeneracy. Therefore the formation of incompressible QH states for $\nu<4$ requires interactions; projecting these onto the lowest LL yield the effective Hamiltonian
 \begin{equation} \label{eq:Hi}
H_i = \frac{1}{2A}\!\!\sum_{\vs{q}\alpha\beta\gamma\delta X X'} \!\!V(q) \; \vs{:}\!\bar{\rho}_{\alpha \beta} (\bar{\vs{q}}_{\alpha \beta}, X) \bar{\rho}_{\gamma \delta} (-\bar{\vs{q}}_{\delta \gamma}, X')\!\vs{:}.\!\!
\end{equation}

Here, $ \vs{:} \ldots  \vs{:} $ denotes normal ordering, $V (q)$ is the Fourier transform of the interaction. In terms of creation operators $c^\dagger_{\kappa, X}$ which create an electron in the LLL orbital $\phi_{\kappa, X}$, the density at wave-vector $q$, projected into the LLL is given by $\bar{\rho} (\vs{q}) = \sum_{\alpha \beta X} F_{\alpha \beta} (\vs{q}, X) \bar{\rho}_{\alpha \beta} (\vs{q}_{\alpha \beta}, X)$, where 
\begin{align}
\bar{\vs{q}}_{\alpha \beta} &= \vs{q} + \vs{K}_{\alpha \beta},  \\ 
\bar{\rho}_{\alpha \beta} (\bar{\vs{q}}_{\alpha \beta} , X) &= F_{\alpha \beta} (\bar{\vs{q}}_{\alpha \beta} , X) c^\dagger_{\kappa, X - \frac{\bar{q}_{y, \alpha \beta}}{2}}  c_{\kappa', X + \frac{\bar{q}_{y, \alpha \beta}}{2}}, \nonumber \\
F_{\alpha \beta} (\vs{q}, X) &= e^{i q_x X} \frac{(4 z'_\alpha z'_{\beta})^{1/4}}{\sqrt{z^*_\alpha + z_{\beta}}} e^{-  \frac{( q_x + i z^*_\alpha q_y) (q_x - i z_{\beta} q_y)}{2 ( z^*_\alpha + z_{\beta})} }\nonumber.% \\
\label{eq:FF}
\end{align}

Besides interactions, we also assume the presence of a long wavelength strain field. This will generically split the valley degeneracy at the single particle level, but at leading order valleys $A, \bar{A}$ are approximately degenerate and split only by $\delta H_\alpha$, as are $B, \bar{B}$.

 \subsection{Hierarchy of terms}

The `form factors' $F_{\alpha\beta}(\vs{q})$ are exponentially sensitive to the momentum difference between the valleys $\alpha, \beta$. At leading order we may thus restrict to  
\begin{align}
H_{i,0} \boldsymbol{:} \; \text{terms in} \; H_i, \; \text{for} \; \alpha = \beta \; \; \gamma = \delta. 
\end{align}

Going to higher order, valley mixing interactions corresponding to near zero \emph{total} momentum transfer in the 2D Brillouin zone are only polynomially suppressed in $a/\ell_B $. Such terms fall into two categories:
 \begin{align}
 H_{i,1} \boldsymbol{:} \; \text{terms in} \; H_i, \; \text{for} \; (\gamma \delta) = (\beta \alpha), \nonumber \\
 H_{i,2} \boldsymbol{:} \; \text{terms in} \; H_i, \; \text{for} \;  (\gamma \delta) = (\bar{\alpha} \bar{\beta}). 
 \end{align}
 
In both $H_{i, 1}, H_{i,2}$ we require $\beta \neq \alpha$, and additionally in $H_{i,2}$, $\beta \neq \bar{\alpha}$. 

For both of the above terms, $\bar{q}_{\delta \gamma} = \bar{q}_{\alpha \beta}$. A transformation $\vs{q} \rightarrow \vs{q} + \vs{K}_{\beta \alpha}$ transfers all dependence on this momentum transfer $\vs{K}_{\beta \alpha}$ into the argument $V(q)$, leading to an overall factor of $\mathcal{O}(a/\ell_B)$ relative to $H_{i,0}$. It is worth noting that these terms may in fact be competitive with $H_{i,0}$ if the Coulomb interaction $V(q)$ is screened on a length scale $l_{\text{sc}} \lesssim 1/K_{AB}, 1/K_{A \bar{B}}$. In this case, the factor $V(q \sim 0) \approx V(q \sim 1/K_{AB}, 1/K_{A\bar{B}}) \sim (e^{2}/\epsilon) \cdot l_{\text{sc}}$. The possibility of tuning the relative strength of these terms with respect to $H_{i,0}$ can allow us to access different FQH states, as we argue below. 

All other terms describe scattering processes with a large net 2D momentum transfer. While these are allowed in principle because of LLL projection, they are exponentially small $\sim e^{- (K \ell_B)^2} \approx e^{- (\ell_B/a)^2}$ and can be safely neglected. 

%For notational convenience, we dub the degree of freedom between two valleys that share the same anisotropy for $\delta H_{\alpha}=0$  (i.e., $X\leftrightarrow \bar{X}$ for $X=A,B$) `pseudospin' and that between such anisotropy pairs ($A\leftrightarrow B$), `isospin'. Domain walls between QHFMs polarized in different valleys are pinned by strain, that we model as a slowly varying valley Zeeman field that couples only  to isospin.
%. We will capture this by a linearly varying potential
 %$H_\Gamma = \Gamma \sum_X r(\alpha) X c^\dagger_{\alpha, X} c_{\alpha, X}$, with $r (A) = r(\bar{A}) = 1$, and $r (B) = r(\bar{B}) = -1$.  
% 
 
 \subsection{Symmetries}
%\textbf{\textit{Symmetries.}} %---}}  
In the elliptical-valley limit, $\delta H_{\alpha}=0$, we can use the approximate the form factors from Eq.~(\ref{eq:FF}). Then, we note that $H_{i,0}$ is invariant under $SU(2)$ rotations within valley pairs $A, \bar{A}$ and $B, \bar{B}$, respectively. 
%This yields a rich symmetry structure~\cite{QHFMOBD} but for our discussion we take $\delta H_\alpha \neq0$ (as is likely case in Bi(111)). However we will approximate the form factors by (\ref{eq:FF}). 
$H_{i,1}$ breaks this $SU(2)$ symmetry but possesses a $[U(1)]^4$ symmetry, namely independent conservation of the electron number $N_\alpha$ in each valley. We can rearrange these into the $4$ $U(1)$ charges introduced above in Eq.~(\ref{eq:charges}). 

Finally, $H_{i,2}$ preserves $\mathcal{Q_\rho}, \mathcal{Q_\sigma}$ and $\mathcal{J_\sigma}$, but breaks the $U(1)$ symmetry associated with $\mathcal{J_\rho}$ as it allows an electron each from valleys $ A, \bar{A}$ to scatter to valleys $B, \bar{B}$; this process changes $\mathcal{J_\rho} \rightarrow \mathcal{J_\rho} \pm 2$. 

%We will use these symmetries below to strongly constrain terms allowed in the low-energy theory of the domain wall.
%We comment here that the model has enhanced symmetry in the elliptical valley limit  $\delta H_{\alpha}=0$, where $H_{i,0}$ is invariant under $SU(2)$ pseudospin rotations. The rich symmetry structure~\cite{QHFMOBD} in this case may lead to additional interesting effects; however here we assume that $\delta H_\alpha \neq0$
 %(as the case in Bi(111)). Therefore we will approximate the form factors by (\ref{eq:FF}). 

%\textbf{\textit{QHFMs at $\nu=1,2$.}} %---}} 
 %We now construct ground states of (\ref{eq:Hi}), ignoring for  the moment intervalley contributions from $H_{i,1,2}$. 
 \subsection{QHFM ground state at $\nu = 2/m$.}
 
 In what follows, we always consider the case where a very weak strain field breaks the $4$-fold valley degeneracy down to two sets of $2$-fold degeneracies. Ignoring the weaker interactions $H_{i,1}, H_{i,2}$, at $\nu = 2$, it is evident that, in the ground state, the LLL in valleys $A, \bar{A}$ will be occupied in the regions these valleys have lower energy than $B, \bar{B}$ valleys due to strain, and vice versa in regions where valleys $B, \bar{B}$ have lower energy. Let us assume the former case for the following discussion. 
 
 In the integer case, an exchange gap $\sim e^2/\epsilon \ell_B$ then separates the LLLs of the occupied $A, \bar{A}$ valleys from the LLLs of the unoccupied $B, \bar{B}$ valleys. Here we note that, at the level of approximation where we only consider $H_{i,0}$, there is an $SU(2)$ rotation symmetry between valleys $A, \bar{A}$, and there is no distinction between filling up LLLs in valleys $A, \bar{A}$ as opposed to any other orthogonal superposition of valleys $A, \bar{A}$. A finite $\delta H_\alpha$ breaks the elliptical valley approximation and will break the degeneracy between valleys $A, \bar{A}$ at the single particle level, but should not alter the ground state occupation of the valleys provided it is weak. 
 
 For fractional quantum Hall states, the situation is more delicate. For instance, at $\nu = 2/3$, two prominent competing states, the Halperin $(3,3,0)$ state, and the pseudospin singlet state can both be viable ground state candidates. (A fully pully polarized $\nu = 2/3$ state is likely to be higher energy as it involves effectively filling two composite fermion LLs.) It is useful to orient our discussion based on numerical findings in the bilayer system. In that system, studied in Ref.~\cite{peterson2015abelian}, when the interlayer distance vanishes interactions are perfectly $SU(2)$ symmetric, and the pseudospin singlet state appears to energetically favored over the $(3,3,0)$ state (see also Ref.~\cite{faugnotwocomponentCF}). This is similar to the situation at hand if we consider only $H_{i,0}$ and neglect $H_{i,1}$ and $\delta H_\alpha$. When the interlayer distance is of the order of, or exceeds the inter-particle distance within a single layer, the Halperin $(3,3,0)$ state is seen to become energetically favorable again. This occurs because repulsion between electrons in the same layer becomes stronger than the repulsion between electrons in difference layers as the interlayer distance is increased. 
 
 (Note that we can ignore in what follows $H_{i,2}$ because it involves all $4$ valleys, and the assumption that the strain field lifts the degeneracy between valleys $A, \bar{A}$ and $B, \bar{B}$, rendering one of these pairs unoccupied.)
 
 In comparison to $H_{i,0}$ which possesses an $SU(2)$ symmetry, $H_{i,1}$, involves an exchange of electrons between valleys $A, \bar{A}$ and thus serves to reduce the net repulsion between electrons in valleys $A$ and $\bar{A}$. In the case when the screening length is short compared to the magnetic length, as noted above, $H_{i,1}$ can become comparable to $H_{i,0}$. We thus anticipate that the Halperin $(3,3,0)$ state may become favorable in the case where the screening is sufficiently strong in this system. Of course, a comprehensive numerical study is required to ascertain which state is favored as a function of the screening length; this is beyond the scope of this work. 
 
 In what follows, we take an agnostic approach and assume that the low energy theory of this system at filling $\nu = 2/m$ can be captured by a Chern-Simons field theory with a $K$ matrix~\cite{wen1991topological,wen1992classification} and charge vector $q$ of the form 
 
 \begin{align}
     K &= \begin{pmatrix} m'  & n \\ n & m' \end{pmatrix}, \; \; q = \begin{pmatrix} 1 \\ 1 \end{pmatrix}
     \label{eq:KmatrixA}
 \end{align}
 
The filling fraction for such a state is given by $\nu = q^{\text{T}} K^{-1} q = 2/(m'+n)$. Here, $m'$ is an odd integer, while $n$ is an even integer, such that $m'+n = m$ is an odd integer. As an example, at $\nu = 2/3$, one may consider two states possessing unique $K$ matrices [unrelated by SL(2,$\mathbb{Z}$) transformations], the pseudospin singlet which corresponds to $m' = 1, n = 2$, while the Halperin $(3,3,0)$ state corresponds to $m' = 3, n = 0$. As we will show, the quantum dimension of the non-Abelian particles realized is $\sqrt{m'-n}$. Thus, at $\nu = 2/3$, only the Halperin $(3,3,0)$ state has the topological degeneracy we seek. Further, for integer quantum Hall states, at $\nu = 2$, there is no topological degeneracy.

\section{Edge model and gapping perturbations}
\label{sec:edge}

We imagine a situation where one half of the system say, $x<0$, has states $A, \bar{A}$ occupied with a K-matrix of the form of Eq.~(\ref{eq:KmatrixA}) while on the other half, $x>0$, valleys $B, \bar{B}$ are occupied with in an FQH state governed by the same K-matrix. One may view the region $x>0$ as a trivial state for the valley charges $A, \bar{A}$, which leads to the presence of gapless chiral edge modes in analogy with the usual setting where edge modes are realized in two dimensional electrons residing in bulk FQH states at the boundaries of their region of confinement. In the hydrodynamic picture of FQH edges, and in the absence of edge reconstruction effects and interactions between edges, these chiral fermionic edge modes can be captured by chiral bosonic fields~\cite{wen1990chiral,wen1995topological,chang2003chiralluttinger,wen2004quantum} $\phi_I (y)$ defined along the edge, with $I$ representing the valleys $A, \bar{A}, B, \bar{B}$ (in that order), and $\rho_I = \frac{1}{2\pi} \partial_y \phi_I $ representing the electronic charge density in valley $I$ at the edge. The density operators are not independent for different $I, J$ in general and satisfy a Kac-Moody algebra with commutation relations 

\begin{align}
    \left[ \phi_I (x), -\frac{1}{2\pi} K_{IJ} \partial_y \phi_J (y) \right] = i \delta (x - y), 
    \label{eq:commrels}
\end{align}

with the K-matrix 
\begin{align}
    K_{IJ} &= \begin{pmatrix} m' & n & 0 & 0 \\ n & m' & 0 & 0 \\ 0 & 0 & -m' & -n \\ 0 & 0 & -n & -m' \end{pmatrix}. 
\end{align}

Including interactions between the edge mode densities, the edge is governed by the action

\begin{align}
    S = - \int \frac{dy dt}{4\pi} \left[ K_{IJ} \partial_t \phi_I \partial_x \phi_J + V_{IJ} \partial_x \phi_{I} \partial_x \phi_{J} \right].
\end{align}

$V_{IJ}$ is a symmetric positive-definite matrix that  depends on microscopic details, but its general form can be gleaned from symmetry arguments. In particular, allowing for arbitrary $V_{IJ}$ that respects the point group symmetries of the Hamiltonian, we can choose 

\begin{align}
    V_{IJ} &= \begin{pmatrix} V_0 + \Gamma & V_1 & V_2 & V_3 \\ V_1 & V_0 + \Gamma & V_3 & V_2 \\ V_2 & V_3 & V_0 + \Gamma & V_1 \\ V_3 & V_2 & V_1 & V_0 + \Gamma \end{pmatrix}. 
\end{align}

Here $V_0, V_1, V_2, V_3$ represent varies density-density interactions between the valley filtered edge modes, while $\Gamma$ is physically dinstict and originates from a strain gradient which ultimately pins the line defect---in particular, this strain gradient serves as an effective valley Zeeman field that splits the degeneracy weakly between valleys $A, \bar{A}$ and valleys $B, \bar{B}$, and varies in a direction orthogonal to the line defect. Alternatively, one can view this as an effective electric field in a direction orthogonal to the line defect, with equal magnitude but pointing in opposite directions for valleys $A, \bar{A}$ and $B, \bar{B}$. The interactions between valley-pairs is chosen to be identical if they are related by a point-group symmetry (for instance, interaction between edge modes $A, \bar{A}$ is equal to that between valleys $B, \bar{B}$). 

The low-energy Hamiltonian of the edge modes is given by

\begin{align}
    H = \int \frac{dy}{4\pi} \; V_{IJ} \partial_x \phi_{I} \partial_x \phi_{J}, 
\end{align}

where the K-matrix is nominally absent, but encodes the commutation relations between the fields as in Eq.~(\ref{eq:commrels}). 

The Hamiltonian can be put into a Luttinger liquid form using an orthogonal transformation with the matrix

\begin{align}
    U &= \frac{1}{2} \begin{pmatrix} 1 & 1 & 1 & 1 \\ 1 & 1 & -1 & -1 \\ 1 & -1 & -1 & 1 \\ 1 & -1 & 1 & -1 \end{pmatrix}.  
\end{align}

yielding two pairs of non-commuting charge $(\phi_\rho, \theta_\rho)$ and pseudospin fields $(\phi_\sigma, \theta_\sigma)$, in that order. The gradients of these fields can be identified with the relevant charges noted in Eq.~(\ref{eq:charges}); for instance, $\mathcal{Q_\rho} = \frac{\nabla \phi_\rho}{2\pi}$ The non-zero commutation relations between these fields are

\begin{align}
    \left[ \phi_\rho (x), \theta_\rho (y) \right] &= -i \frac{2\pi}{m'+n} \Theta (y-x) \nonumber \\
    \left[ \phi_\sigma (x), \theta_\sigma (y) \right] &= -i \frac{2\pi}{m'-n} \Theta (y-x).
    \label{eq:commrels}
\end{align}

where $\Theta(x)$ is the Heaviside theta-function in $x$. Finally, the Luttinger Hamiltonian is given by

\begin{align}
    H = \sum_{\nu = \rho, \sigma} \int \frac{dy}{2\pi} \left[ v_\nu K_\nu (\nabla \theta_\nu)^2 + \frac{v_\nu}{K_\nu} (\nabla \phi_\nu)^2 \right]. 
\end{align}

with appropriate Luttinger parameters $K_\nu$ and velocities $v_\nu$ for $\nu = \rho, \sigma$. 

We now make note of another physical constraint that imposes a condition on the parameters $V_i$. Specifically, in the absence of a pinning potential provided by $\Gamma \neq 0$, we can expect the line defect to able to translate in the orthogonal direction at no cost~\cite{mitragirvin,agarwal2019topology}. This translation can be viewed as the transfer of charge in regions with occupation of valleys $A, \bar{A}$ to regions with occupation of valleys $B, \bar{B}$ (or vice versa), and concomitantly involves a change in the charge density $\mathcal{J_\rho} \rightarrow \mathcal{J_\rho} + \delta$, for arbitrary $\delta$, while leaving other charge densities invariant. Thus, for $\Gamma = 0$, the coefficient of the term $(\nabla \theta_\rho)^2$ should vanish; this implies $V_0 + V_1 = V_2 + V_3$. For finite $\Gamma$, the Luttinger parameter of the charge channel is then given by

\be
    K_\rho = \sqrt{\frac{\Gamma}{2 (V_0 + V_1) + \Gamma}} \sim \sqrt{\Gamma}
    \label{eq:Krho}
\ee

which goes to zero in the limit of a vanishing strain gradient $\Gamma \rightarrow 0$. This in particular implies that fluctuations in $\phi_\rho$ are rather costly; in the language of the renormalization group, it is equivalent to stating that a sine-Gordon term in $\phi_\rho$ can easily pin the field to a given minimum. We will next show that such a term naturally arises in this system and the charge channel will be generically gapped. 

The Luttinger parameter $K_\sigma \rightarrow 1$ for the pseudospin channel in the limit $V_2 \rightarrow V_3$, that is, when the interaction strength between edge modes of valley $A$ interacts equally strongly with edge modes of valley $B$ and valley $\bar{B}$. This can also be seen more directly by noting that substituting $A \leftrightarrow \bar{A}$ while keeping $B, \bar{B}$ fixed sends $K_\sigma \rightarrow 1/K_\sigma$. In what follows, we will also consider perturbations that can pin fields $\phi_\sigma, \theta_\sigma$; there is no choice of $K_\sigma$ that can simultaneously guarantee the relevance of these two perturbations. We assume, as in previous works, however that the pinning perturbation is strong enough to have these fields pinned coherently over the length of the region over which one of these perturbations is in effect. 

\subsection{Electronic operators and gapping perturbations}

We now consider gapping perturbations. To compute these, we first identify the appropriate quasiparticle and electronic operators at the edge. Quasiparticles operators are specified with a vector $l_I$ of integers, with
\be
\psi_{\text{qp}, l_I} (y) \sim e^{i l_I \phi_I (y)}
\label{eq:qpop}
\ee
which change the charge density in mode $I$ as 
\be 
\delta \rho_I (x) = - \delta (x-y) \left( l^{\text{T}} K^{-1} \right)_I = - \delta (x-y) L_I
\ee
where we have and set $l_I = K_{IJ} L_J$. The above can be ascertained by the commutation relation $\left[ \frac{\nabla \phi_I (x)}{2\pi}, e^{i l_I \phi_I (y)} \right] = - \left( l^{\text{T}} K^{-1} \right)_I \delta (x-y) e^{i l_I \phi_I (y)}$. 

Electronic annihilation operators involve a net decrease of charge by $1$. An electron annihilation operator that annihilates one electron in valley $I$ can be found by setting $L_J = \delta_{IJ}$, and computing the corresponding $l_J$. In particular, the electronic annihilation operators in the different valleys are given by 

\begin{align}
    \psi_{ \text{el}, A} (\vs{r}) &\sim e^{i ( m' \phi_A + n \phi_{\bar{A}}) - i\vs{K}_A \cdot \vs{r}}, \nonumber \\
    \psi_{\text{el}, \bar{A}} (\vs{r}) &\sim e^{i ( n \phi_A + m' \phi_{\bar{A}}) - i\vs{K}_{\bar{A}}, \cdot \vs{r}} \nonumber \\
    \psi_{\text{el}, B} (\vs{r}) &\sim e^{-i ( m' \phi_B + n \phi_{\bar{B}}) - i\vs{K}_B \cdot \vs{r}}, \nonumber \\
    \psi_{\text{el}, \bar{B}} (\vs{r}) &\sim e^{-i ( n \phi_A + m' \phi_{\bar{B}}) - i\vs{K}_{\bar{B}} \cdot \vs{r}}. 
    \label{eq:electronoperators}
\end{align}

Here we have also introduced the position-dependent phase owing to the fact that these electronic operators are sourced from valleys centered at momenta $\vs{K}_I$. Alternatively, one can imagine solving for electronic operators in a system with a valley centered at the $\Gamma$ point as usual; then a solution for the electronic operators at the edge for the case when valleys are centered at momenta $\vs{K}_I$ are obtained simply by a gauge transformation with appends the appropriate momentum phase factors to these operators. 

These momenta have an important effect on interactions that remain relevant in this system.  One perturbation $V_{\phi_\rho}$ that arises from the local scattering of electrons between all $4$ valleys is given by 

\begin{align}
    V_{\phi_\rho} \sim \psi^{\phantom{\dagger}}_{\text{el},A} \psi^{\phantom{\dagger}}_{\text{el},\bar{A}} \psi^\dagger_{\text{el},B} \psi^\dagger_{\text{el},\bar{B}} (\vs{r}) &\sim e^{i(m+n) (\phi_A + \phi_{\bar{A}} + \phi_{B} + \phi_{\bar{B}})} \nonumber \\
    & \sim e^{i 2 (m'+n) \phi_\rho}
    \label{eq:Vrho}
\end{align}

where we used the fact that $\vs{K}_A + \vs{K}_{\bar{A}} = 0$, $\vs{K}_B + \vs{K}_{\bar{B}} = 0$. All other such interaction terms involving the $4$ valleys involve oscillations at short wavelengths and are in general irrelevant. We note that the term noted above can be viewed as a backscattering term that takes electrons from valleys $A, \bar{A}$ and converts them to electrons in valleys $B, \bar{B}$ and can gap the charge channel by pinning the field $\phi_\rho$. (Note that the domain wall itself can provide a momentum kick $\sim 1/l_{\text{DW}}$ orthogonal to it, where $l_{\text{DW}}$ describes its thickness. In the limit where the valleys are strongly anisotropic, when $\lambda \gg 1$ or $\lambda \ll 1$, this length shortens and can become of the order of the magnetic length~\cite{abanin2010nematic}, $\ell_B$, which is nevertheless smooth on the atomic length scale; thus the domain wall by itself cannot provide for the momentum mismatch between valleys.)

Note also that thus far we have ignored noting Klein factors which are necessary to capture the anti-commutation of electronic operators. There are two interaction events that involve the same net transfer of electrons between valleys. In particular, interaction events sending i) $A \rightarrow B, \bar{A} \rightarrow \bar{B}$ with an amplitude, say $V_{\text{i}}$, or ii) $A \rightarrow \bar{B}, \bar{A} \rightarrow B$, with an amplitude, say $V_{\text{ii}}$ have the same form as in Eq.~(\ref{eq:Vrho}) but come with a relative minus sign due to the appropriate positioning of the electronic operators. The amplitudes of these processes are generically different but real---the reverse process should have the amplitudes $V^*_{\text{i,ii}}$ by hermitian conjugation, but mirror reflection about the symmetric axis between valleys $A, B$ achieves the same reversal. If the valleys are related by a $\pi/2$ rotation, then the processes should have the same amplitude and will cancel each other out; thus we assume, as is true for Si(111), and Bi(111) surfaces, that 2DBZ does not possess a $\pi/2$ rotation symmetry. Then, the two amplitudes $V_{\text{i,ii}}$ are inequivalent and the gapping perturbation then has the form 

\begin{align}
    V_{\phi_\rho} &\sim \abs{A_{\phi_\rho}} \cos \left( 2 (m'+n) \phi_\rho\right) 
    \label{eq:chargepert}
\end{align}

for some amplitude $A_{\phi_\rho}$. Since there is an exceedingly large cost to fluctuations in the field $\phi_\rho$ in the limit of vanishing strain gradient $\Gamma$ [see Eq.~(\ref{eq:Krho})], we expect that the charge channel is always gapped by such a perturbation. 

\subsection{Superlattice aided gapping perturbations}

In order for us to realize the setup of Fig.~\ref{fig:prototype}, we need to realize perturbations that pin the non-commuting fields $\phi_\sigma, \theta_\sigma$ in alternating regions along the line defect. We will see that these processes can be realized if a superlattice is imposed on the line defect, and which provides a momentum $\vs{q}_{\text{SL}} = 2 \vs{K}_{AB}$ or $2 \vs{K}_{A \bar{B}}$, to pin the fields $\phi_\sigma$ and $\theta_\sigma$ respectively. In particular, we imagine this superlattice coupling to the underlying FQH surface as an oscillatory potential at the appropriate wavevector. 

Let us first consider the processes that pin the field $\phi_\sigma$. As an example, consider the process where due to the superlattice potential, an electron in valley $A$ first scatters out of the valley mimumum, absorbing momentum $\vs{q}_{\text{SL}} = 2\vs{K}_{AB}$; see Fig.~\ref{fig:figbands} (d) for an illustration. Subsequently, this off-shell electron interacts with another electron in valley $\bar{B}$ to relax into valley $B$, with the other electron transferring to valley $\bar{A}$. Overall, this process results in two electrons from valleys $A, \bar{B}$ transferring to states in valleys $B, \bar{A}$, with a net momentum transfer of $\vs{q}_{\text{SL}} = 2 \vs{K}_{AB}$ in the 2DBZ. Such a process will be smaller than the interaction induced gapping perturbation $V_\rho$ in the charge channel by an additional factor of $V_{\text{SL}}/W$, where $V_{\text{SL}}$ is the amplitude of the superlattice potential, and $W$ is of the order of the bandwidth of the bands associated with the valleys in which the FQH states are realized. From the defintion of charges in Eq.~(\ref{eq:charges}), it is easy to see that this process results in the change $\mathcal{J_\sigma} \rightarrow \mathcal{J_\sigma} \pm 2$ while leaving all other charges unchanged; using the commutation relation between the fields and charge densities, it is straightforward to show note this is realized by a term $\sim \cos (2 (m'-n) \phi_\sigma )$, which in turn pins $\phi_\sigma$. 

We now consider all such processes that give rise to a perturbation that pins the field $\phi_\sigma$. There are $3$ unique processes that involve an electron from valley $A$ absorbing momentum $2 \vs{K}_{AB}$. Specifically, after absorbing this momentum, there are $3$ unique interaction processes of generically different amplitudes that can take place. We enumerate these in terms of the scattering of the electron still residing in the valley (while the now off-shell electron from valley $A$ scatters by an equal and opposite momentum). These are i) an electron from valley $\bar{B}$ scatters to valley $\bar{A}$, i') an electron from valley $\bar{B}$ scatters to valley $B$, and ii) another electron from valley $A$ scatters to valley $B$. The first two processes involve the same change in the charge quantum numbers and thus have the same form; they are illustrated by blue and green lines in Fig.~\ref{fig:figbands} (d). We will denote them with a net amplitude $V_{\text{i}}$. The last process will be denoted with a amplitude $V_{\text{ii}}$. 

In addition to the above $3$ processes, we can also enumerate processes that involve an electron from valley $\bar{B}$ absorbing the superlattice momentum and then undergoing an interaction induced scattering. These processes are related to the above by $\pi$-rotation and subsequent reflection about the symmetry axis and thus come with the same amplitude; although they can involve different changes to the charge quantum numbers. Finally, all these processes of course appear with the time-reversed  reverse process involving absorbing net momentum $-2 K_{AB}$ from the superlattice. Their amplitude is related to the former by complex conjugation. 

Using the above symmetry considerations, all the relevant processes add up to yield

\begin{align}
    V_{\phi_\sigma} & \sim 2 \abs{V_{\text{i}}} \cos \left( 2 (m'-n) \phi_\sigma + \varphi_{\text{i}} \right) \nonumber \\
    &+ 4 \abs{V_{\text{ii}}} \cos \left( 2(m'+n) \phi_\rho \right) \cos \left(2 (m'-n) \phi_\sigma  + \varphi_{\text{ii}} \right), \nonumber \\
    & \sim \abs{A_{\phi_\sigma}} \cos \left( 2 (m'-n) \phi_\sigma + \varphi_{\phi_\sigma} \right). 
\end{align}

where $\varphi_{\text{i, ii}}$ are undetermined phases that depend on microscopic details. We note that the second term involving $\phi_\rho$ may be viewed analogously to the first term as it is reasonable to assume that $\phi_\rho$ is pinned by the more relevant perturbation in Eq.~(\ref{eq:chargepert}). In this way, we realize an appropriate gapping term that pins the field $\phi_\sigma$. 

In analogy to the above, we can imagine other sections of the line defect where a superlattice providing the wavevector $\vs{q}_{\text{SL}} = \vs{K}_{A \bar{B}}$ is present. The above analysis can be repeated to yield a perturbation that pins the field $\theta_\sigma$---

\begin{align}
    V_{\theta_\sigma} \sim \abs{A_{\theta_\sigma}} \cos \left( 2 (m'-n) \theta_\sigma + \varphi_{\theta_\sigma} \right)
\end{align}

\subsection{Luttinger model of the line defect}

We can thus summarize the Luttinger model of the line defect, in our scheme, as follows

\begin{align}
    H \sum_{\nu = \rho, \sigma} & \int \frac{dy}{2\pi} \left[ v_\nu K_\nu (\nabla \theta_\nu)^2 + \frac{v_\nu}{K_\nu} (\nabla \phi_\nu)^2 \right] \\ \nonumber
    + & \int \frac{dy}{2\pi} V_\rho \cos \left( 2 (m'+n) \phi_\rho + \varphi_\rho \right) \nonumber \\
    + & \int \frac{dy}{2\pi} V_{\phi_\sigma} (y) \cos \left( 2(m'-n) \phi_\sigma + \varphi_{\phi_\sigma} \right) \nonumber \\
    + & \int \frac{dy}{2\pi} V_{\theta_\sigma} (y) \cos \left( 2(m'-n) \theta_\sigma + \varphi_{\theta_\sigma} \right) . 
\end{align}

As noted above, we assume that the field $\phi_\rho$ in the charge channel is pinned everywhere along the line defect by a relevant perturbation (since we can make $K_\rho$ arbitrarily small). The two other gapping perturbations have an amplitude $V_{\phi_\sigma}, V_{\theta_\sigma}$ that is alternated in magnitude along the line defect, with the help of orthogonal superlattices, and which then serve to pin two non-commuting fields $\phi_\sigma, \theta_\sigma$ in the pseudospin channel. This gives us the topological ground state degeneracy as discussed in Sec.~\ref{sec:summary}.

\section{Ground state subspace, non-Abelian zero modes and braiding scheme}
\label{sec:nonabelian}

We now discuss more carefully the ground state subspace of this system, provide explicit constructions of the non-Abeliab zero modes, and analyze how they can be braided. We will see that the quantum dimension of the non-Abelian quasiparticles realized in this system scales as $\sqrt{\abs{m'-n}}$. Thus, for instance, at $\nu = 2/3$, the pseudospin singlet state with $m' = 1, n = 2$ does not host a topological degeneracy while the Halperin $(3,3,0)$ state with $m' = 3, n = 0$ hosts non-Abelian quasiparticles with quantum dimension $\sqrt{3}$.  

\subsection{Ground state subspace}

Given that the charge channel is gapped all along the line defect, we do not invoke it except to note that all low energy operators must operate trivially in that channel. 

In the pseudospin channel, we assume that the perturbations $V_{\phi_\sigma}$ and $V_{\theta_\sigma}$ are strong enough to gap out the entire length of the region where such a perturbation applies and pins $\phi_\sigma, \theta_\sigma$, respectively. Let us consider a circular line defect divided into $2N$ equal segments of length $L$ with interfaces labeled by $j = 1, \dots, 2N$, and fields $\phi_\sigma, \theta_\sigma$ taking values $\phi_i, \theta_{i+1}$ between interfaces $2i-1, 2i$ and $2i, 2i+1$, respectively, for $i = 1, \dots, N$.  

One can view segments between consecutive regions where $\phi_\sigma = \phi_{i}, \phi_{i+1}$ is pinned, as having fixed charge $\mathcal{Q}_\sigma = Q_i \equiv \left( \phi_{i+1} - \phi_i \right)/2\pi $. The perturbation $V_{\theta_\sigma}$ in this region is such that it translates the charge $Q_i \rightarrow Q_i \pm 2$. (One can verify this fact using the appropriate commutation relations.) Since it is the relevant gapping perturbation in this region, it coherently couples all states with charge $Q_\sigma$ separated by units of $2$, and as a result, the ground state can be labeled by a charge $Q_i$ that is defined modulo $2$. (If this charge was the physical charge, we would be considering here a usual superconductor whose two low energy states are identified by the charge parity defined modulo $2$.) The segments next to this region are insulators of $\mathcal{Q}_\sigma$ charge as they have gapping perturbations that pin the value of $\phi_\sigma$ to a constant. Thus, in order to change the value $Q_i$, ($Q_\sigma$) charge must traverse an insulating barrier of length $L$ and arrive from regions labeled by $Q_{i-1}, Q_{i+1}$. This process is exponentially suppressed as $e^{-L/\xi}$, where $\xi$ is the typical correlation length of $\theta_\sigma$ in segments where $\phi_\sigma$ is pinned. As a result, states labeled with different values of $Q_{i}$ have an energy splitting that is exponentially small in the size $L$ of the segments. 

Since fields $\phi_\sigma, \theta_\sigma$ cannot be simultaneously specified as they do not commute with one another, the ground state is uniquely identified by specifying $\phi_\sigma = \phi_i$, on $i = 1,\dots, N-1$ segments out of a total of $N$ segments where $\phi_\sigma$ is pinned. We have used the fact that the total charge on the line defect, $\sum_i Q_i$ is conserved to note that there are only $N-1$ independent $\phi_i$. 

It now remains to determine which values of $Q_i$ correspond to unique and physical ground states. The physical operators in our system are the quasiparticle operators noted in Eq.~(\ref{eq:qpop}) and specificed by an integer $l_I$. Using the commutation relations between such quasiparticle operators and the charges, we can see that the quasiparticle operator that minimally changes $Q_i$ while leaving the charges $\mathcal{Q_\rho, J_\rho}$ unchanged, changes $Q_i \rightarrow Q_i \pm 1/(m'-n)$. One example of such a quasiparticle operator is $\sim e^{i \phi_A - i \phi_{\bar{A}}} = e^{i \phi_\sigma + i \theta_\sigma}$ inserted at one of the two interfaces adjacent to the segment with charge $Q_i$. 

Now, since $Q_i$ is identified modulo $2$, we may expect $2(m'-n)$ distinct values for each independent $Q_i$. In fact, as we now show, there are only $m'-n$ distinct minima. In the basis where the charges $Q_i$ take well defined values, the operator $e^{2\pi i (m'-n) Q_i} \equiv \mathbb{1}$, given the quantization of the charge in units of $1/(m'-n)$. However, despite being clearly $\mathbb{1}$ in one basis, one can also confirm that this operator changes $J_{i\pm 1} \rightarrow J_{i\pm 1} \mp 1$. We can reverse the argument to come to the conclusion that the state with charges $\dots, Q_i, \dots ,Q_j, \dots$ must be identical to that with $\dots, Q_i \pm 1, \dots ,Q_j \mp 1, \dots$, for any pair of $Q_i, Q_j$. These symmetries, along with the total charge being conserved modulo $2$, can be used to bring all independent $Q_i \forall i \in 1, \dots, N-1$ to the range $[0,1)$. In effect, we see that $Q_i$ can be identified modulo $1$, leaving us with $m'-n$ distinct values for each independent $Q_i$, and ground states can be labeled with the expectation value of the operator $e^{i 2\pi Q_i}$. In total, the ground state subspace is $(m'-n)^{N-1}$ fold degenerate.

Another useful set of operators acting on the ground state subspace can be composed of the conjugate charges $\mathcal{J_\sigma}$. Defining $e^{i 2 \pi J_i} = e^{i(\theta_{i+1} - \theta_{i})}$, it is easy to check the commutation relations

\begin{align}
    Q_i e^{i 2\pi J_i} &= e^{i 2\pi J_i} \left( Q_i - \frac{1}{m'-n} \right) \nonumber \\
    Q_i e^{i 2\pi J_{i+1}} &= e^{i 2\pi J_{i+1}} \left( Q_i + \frac{1}{m'-n} \right)
\end{align}

which imply that operators $e^{i 2\pi J_i} \left( e^{i 2\pi J_{i+1}} \right)$ decrease (increase) $Q_i$ by $1/(m'-n)$. We note for instance $e^{i 2\pi J_2}$ can be constructed from the physical quasiparticle operator $\sim e^{i \phi_A - i \phi_{\bar{A}}} = e^{i \phi_\sigma + i \theta_\sigma}$ and its hermitian conjugate applied on interfaces $3$ and $4$ respectively and projecting them on to the ground state subspace. Thus these operators can be physically realized (as we discuss more later). 

One can also identify the ground state degeneracy, $(m'-n)^{N-1}$, by finding the minimal set of pairs of non-commuting operators that commute among each other, in an analogous fashion to the discussion of Wilson loops in Fig.~\ref{fig:wilsonloop}. In particular, we can identify the following pairs

\begin{align}
    \left( \prod_{j \le i} e^{i 2\pi J_j} \right) e^{i 2\pi Q_k} = e^{i \frac{2\pi \delta_{ik}}{m'-n} } e^{i 2\pi Q_k} \left( \prod_{j \le i} e^{i 2\pi J_j} \right)
    \label{eq:noncommops}
\end{align}

For each pair of such non-commuting operators, the minimal irreducible representation has dimension $m'-n$. These operators are in fact analogous to the Wilson loop operators identified in Fig.~(\ref{fig:wilsonloop})---integrating out the time-component $a_{t,I}$ in the low-energy Chern Simons theory yields the flux condition $\epsilon_{ij} \partial_i a_{j,I} = 0$, or $a_{j,I} = \delta_j \phi_I$; the line integral of $\vs{a}_{I}$ terminating at the line-defect then yields operators $\sim e^{i (\phi_i - \phi_{i+1})}$ equivalent to the opertors identified in Eq.~(\ref{eq:noncommops}). This guarantees the ground state degeneracy of $(m'-n)^{N-1}$.

\subsection{Non-Abelian quasiparticles}

The above physical operators that toggle the ground state subspace can be constructed from pairs of non-Abelian zero modes residing at the interface between segments gapped out by different perturbations. To define these operators, we first need to consider the complete ground state subspace including topological sectors of different total charges $\int dy \mathcal{Q_\sigma} (y), \int dy \mathcal{J_\sigma} (y)$ in the pesudospin sector, but fixed total charges in the charge sector.  

Suppose, without loss of generality, we fix $\int dy \mathcal{Q_\rho} (y) = \int dy \mathcal{J_\rho} (y) = 0$. This implies $\mathcal{N}_{A} = - \mathcal{N}_{\bar{A}}, \mathcal{N}_{B} = - \mathcal{N}_{\bar{B}}$, where $\mathcal{N}_\kappa = \int dy \rho_\kappa (y)$ is the total charge in valley $\kappa$. Due to the gapping perturbations, the total pseudospin charges $\mathcal{N}_\sigma \equiv \int dy \mathcal{Q_\sigma} (y) = \mathcal{N}_A + \mathcal{N}_B, \; \mathcal{I}_\sigma \equiv \int dy \mathcal{J_\sigma} = \mathcal{N}_A - \mathcal{N}_B$ are only defined modulo $2$. Further, as mentioned above, these total charges can be increased in units of $1/(m'-n)$ independently using successive applications of quasiparticle operators $\psi^\dagger_A \psi_{\bar{A}}$ and $\psi^\dagger_B \psi_{\bar{B}}$ which do not affect the other charges. With the additional symmetry---$\mathcal{N}_A, \mathcal{N}_B \rightarrow \mathcal{N}_A \pm 1, \mathcal{N}_B \pm 1$ (from the identification of the charges modulo $2$), the number of topologically distinct total charge sectors are given by $2(m'-n)^2$ and can be spanned by the choice $(m'-n) \mathcal{N_\sigma} = 0, 1, \dots, m'-n-1$, and $(m'-n) \mathcal{I_\sigma} = 0, 1, \dots, 2(m'-n)-1$.  

Thus, we can label ground states as $\ket{q_1, \dots, q_N, \mathscr{J}}$ with $\avg{e^{i 2\pi Q_i}} \equiv e^{i 2\pi q_i/(m'-n)}$ for $i \in 1, \dots, N$ in the relevant segments with integer $q_i$ defined modulo $m'-n$, and $\mathscr{J}$ as the (integer) eigenvalue of $(m'-n) \mathcal{I_\sigma}$, defined modulo $2(m'-n)$. 

In order to construct the non-Abelian zero modes, we now define the following operators

\begin{align}
    &e^{i \hat{\theta}_{j+1}} \ket{q_1,..., q_j,..., q_{N}, \mathscr{J}} = \ket{q_1, ..., q_j + 1,..., q_{N}, \mathscr{J}}, \nonumber \\
    &e^{i 2\pi \hat{Q}_j} \ket{q_1, \dots, q_j, \dots, q_{N}, \mathscr{J}} = \nonumber \\
    & \hspace{2 cm } e^{i 2\pi (m'-n) q_j} \ket{q_1, \dots, q_j, \dotsm q_{N}, \mathscr{J}}, \nonumber \\
    & \hat{T}_{\mathscr{J}} \ket{q_1, \dots, q_j, \dots, q_{N}, \mathscr{J}} = \nonumber \\
    & \hspace{ 2 cm} \ket{q_1, \dots, q_j + 1, \dotsm q_{N}, \mathscr{J} + 1}, \nonumber \\
    \label{eq:gsopdef}
\end{align}

%where we have labeled ground states Note that here we are considering the larger Hilbert space of ground states that live in different topological sectors of the total charge $q \equiv \sum_i q_i$ modulo $m'-n$ and total current $\mathscr{J}$ defined modulo $2(m'-n)$, though we will eventually work in the subspace where these are fixed. Note that 
%
%Here we are considering the larger ground state subspace where the total charge $q \equiv \sum_i q_i$ modulo $m'-n$, and total current $\mathscr{J}$ defined modulo $2(m'-n)$ are undetermined, though we will eventually work in a subspace where these are fixed. 

The operator $e^{i \hat{\theta}_{j+1}}$ defined above sends $q_j \rightarrow q_j + 1$, without altering $\mathscr{J}$, while the operator $T_{\mathscr{J}}$ sends the total conjugate current $\mathscr{J} \rightarrow \mathscr{J} + 1$ without altering the total charge $q$. One can easily surmise that it is not physically possibly to minimally change $\mathcal{Q_\sigma}$ without simultaneously affecting at least one other charge. If we want to keep the charge channel undisturbed, the remaining choice is that we simultaneously change $\mathcal{J_\sigma}$. Thus, operators $e^{i \hat{\theta}_j}$ and $\hat{T}_\mathscr{J}$ are unphysical by themselves, but a combination of them can be.

In particular, one can define two flavors of non-Abelian zero modes that reside at the interaface between alternately gapped segments as 
\begin{align}
    \chi_{2j, \sigma} &= \hat{T}^\sigma_{\mathscr{J}} e^{i \hat{\theta}_{j+1}} \prod_{i < j} e^{i2\pi \sigma Q_i} \nonumber \\
    \chi_{2j - 1, \sigma} &= \hat{T}^\sigma_{\mathscr{J}} e^{i \hat{\theta}_{j}} \prod_{i < j} e^{i2\pi \sigma Q_i}. \\
    \label{eq:gsops}
\end{align}

with $\sigma = +1 (A), -1 (B)$. Using the relations in Eq.~(\ref{eq:gsopdef}), we can observe that these operators satisfy the non-trivial commutation relations for $i<j$

\begin{align}
\chi_{i,\sigma} \chi_{j, A} = \chi_{j, \uparrow} \chi_{i, \sigma} e^{-i 2\pi /(m'-n)} \nonumber \\
\chi_{i,\sigma} \chi_{j, B} = \chi_{j, \downarrow} \chi_{i, \sigma} e^{i 2\pi /(m'-n)} \nonumber \\
\label{eq:nonabelian}
\end{align}

It is not difficult to show that the relations Eq.~(\ref{eq:gsopdef},\ref{eq:nonabelian}) are satisfied by the operators $\psi^\dagger_{\text{qp},A} \psi_{\text{qp}, \bar{A}} (\vs{r}) \sim e^{i (\theta_\sigma + \phi_\sigma)} (\vs{r})$  (thus, analogous to $\chi_A$) and $\psi^\dagger_{\text{qp},B} \psi_{\text{qp}, \bar{B}} (\vs{r}) \sim e^{i (\theta_\sigma - \phi_\sigma)} (\vs{r})$ (thus, analogous to $\chi_B$) if we consider their action at the interface between alternately gapped segments. Thus, we can identify the non-Abelian zero modes as arising from the projection of these operators onto the ground state manifold. 

We can further surmise that these operators lie at the interface between two gapped segments. In particular, we are interested in matrix elements of these quasiparticle operators sandwiched between two states in the ground state subspace. One can show that these matrix elements of these operators (see Ref.~\cite{lindner2012fractionalizing}) fall off exponentially in the distance $r$ from the interface. This essentially follows from the exponential decay of correlations of one of $\avg{e^{i \theta_\sigma (r) - i \theta_\sigma (0)}}$ or $\avg{e^{i \phi_\sigma (r) - i \phi_\sigma (0) }}$ on either side of the interface. 

The commutation relations, Eq.~(\ref{eq:nonabelian}) are different in one major respect from the results of Ref.~\cite{lindner2012fractionalizing}. For integer QH states, Ref.~\cite{lindner2012fractionalizing} obtained the usual fermionic anti-commutation relations, showing that the modes thus obtained correspond to fractionalized Majorana zero modes. In this case, we find that the edge modes commute in the integer case and reduce to bosons and we do not expect any topological degeneracy. Thus, being in a fractional quantum Hall state is essential to this proposal. Furthermore, it is necessary that $m'-n \neq 1$; thus, a pseudospin singlet at $\nu = 2/3$, for instance, is not amenable to our proposal for realizing a topological qubit in this system. 

Next, it is important to understand how these zero modes couple when brought into physical proximity with one another. Since these quasiparticle operators are sourced from FQH states in valleys that are not centered at zero momentum, they carry (fractional) momenta---the momentum carried by a quasiparticle in each mode $I$ can be read out by using appropriate combinations of the electron operators in Eq.~(\ref{eq:electronoperators}). In particular, 

\begin{align}
    \psi_{\text{qp},\kappa} (\vs{r}) \sim e^{i r_\kappa \left[ \phi_\kappa (\vs{r}) -i\vs{K}_\kappa \cdot \vs{r} / (m'-n) \right]}, 
\end{align}

where $r_\kappa = +1 (-1)$ for valleys $\kappa = A, \bar{A} (B, \bar{B})$. 

Now, when zero modes are brought together, interactions can scatter quasiparticles between the edge modes, and are relevant as long as the total momentum in the 2DBZ is conserved. (Note that we are assuming that when the zero modes are brought close, the line defect remains smooth at least on the scale of the magnetic length, which is further assumed to be much greater than the atomic scale $K_{AB}^{-1}$.) Consequently, we only obtain terms such as $\chi^\dagger_{i, A} \chi_{j, A}, \chi^\dagger_{i, B} \chi_{j, B}$ when zero modes at the $i,j$ interfaces are brought close to one another. In fact, as shown in the prototype we consider in Fig.~(\ref{fig:prototype}), if these non-Abelian zero modes couple to one another through a region where only valleys $A, \bar{A}$ are occupied, by proximity, the interaction $\chi^\dagger_{i,B} \chi_{j, B}$ is exponentially weaker and can be neglected in comparison to the coupling $\chi^\dagger_{i,A} \chi_{j, A}$. We assume this going forwards. (Note that a similar assumption is also suggested and in fact required for braiding to be possible in Ref.~\cite{lindner2012fractionalizing}.)

Finally, we note there are $4$ more flavors of charge neutral quasiparticle operators in our system, which are obtained by projecting $\sim \psi^\dagger_{\text{qp}, \kappa} \psi_{\text{qp}, \kappa'}$ for $\kappa \neq \kappa'$ and involving one of each of $A, \bar{A}$ and $B, \bar{B}$ valleys. In the Luttinger language, these correspond to operators $e^{\pm i\phi_\sigma \pm i \phi_\rho}, e^{\pm i \theta_\sigma \pm i \phi_\rho}$. %One can construct operators similarly to above, but to do so, it is necessary to enlarge the ground state subspace noted in Eq.~(\ref{eq:gsopdef}) to include 
Such operators entail changes to the total current in the charge channel, $\mathcal{I_\rho} = \int dy \mathcal{J_\rho} (y)  \rightarrow \mathcal{I_\rho} \pm 1/(m'+n)$ and can amount to additional zero modes only if the charge sector also has topological degeneracy, for instance, if the whole setup of Fig.~\ref{fig:prototype} were to be placed on a torus. For the planar geometry we consider, there is generally a unique ground state in the charge sector, and a change in the current $\mathcal{I_\rho}$ involves creating a quasiparticle or qausihole excitation in the central region which costs finite energy, corresponding to the bulk gap. Thus, the operators $\chi_{i, A (B)}$ are the only zero modes in this setup. 

%We discuss these operators in more detail in Appendix~\ref{sec:flavors}. 
%We note here that such operators can also lead to additional terms in the Hamiltonian when two interfaces are brought together, by pairing up. However, again they can be assumed to be weaker as they involve valley charges $B, \bar{B}$ and thus do not change the arguments that follow.

\subsection{Braiding scheme}

Finally we detail a scheme to braid these zero modes. The braiding scheme is based off of the usual scheme employed to braid Majoranas in T-junction setups~\cite{alicea2011non,halperin2012adiabatic}. In general, one spawns a pair of zero modes, say $2,3$ from the vacuum which are then made to interact with modes $4,5$ to enable their exchange or braid. In particular, first, modes $3,4$ are brought together, effectively `copying' zero mode $4$ on to zero mode $2$, one of the newly spawned zero modes. Subsequently, modes $3,5$ are brought together, copying zero mode $5$ on to mode $4$, achieving `half' of the braiding process. Finally, modes $2,3$ are again brought together to be resent into the vacuum state; this transfers information from zero mode $2$ to mode $5$. This completes the braiding process by effectively exchanging zero modes $4,5$ through an intermediary set of zero modes $2,3$. 

In the prototype topological qubit considered here, as shown in Fig.~\ref{fig:prototype}, we can achieve a resource vacuum by simply pinching interfaces $2,3$ towards the center of the circle to shrink the size of the line defect between these two interfaces; see Fig.~\ref{fig:prototype} (b). This strongly couples modes $2,3$ into a `vacuum' state, and leads to one large segment between interfaces $1,4$ where the field $\phi_\sigma$ is pinned. To pull these modes back from vacuum, we simply separate the zero modes by using strain to expand the region with valleys $A, \bar{A}$ are occupied between interfaces $2,3$. 

Coupling between zero modes at interfaces $i,j$ can be achieved in an analogous fashion. Since the zero modes largely couple to one another through a region where valleys $A, \bar{A}$ are occupied, we anticipate that by proximity, the coupling $\chi^\dagger_{i,A} \chi_{j, A}$ involving edge modes from these valleys significantly dominates the coupling $\chi^\dagger_{i, B} \chi_{j, B}$. As we discuss next, this is an important requirement in ensuring a topologically protected degeneracy of the ground state manifold throughout the braiding process and is thus necessary to prevent non-universal dynamical phases from corrupting the braiding process. We assume this to be the case and now determine the unitary transformation realized as a consequence of these machinations. 

\subsubsection{Hamiltonians realized and symmetries protecting ground state degeneracy}

We can think of the braiding process as being broken into three stages with the following Hamiltonians---

\begin{align}
    H_{\text{I}} &= \lambda (t) H_{34} + (1-\lambda (t)) H_{23} , \; \; \;  t \in \left( t_{\text{I},i}, t_{\text{I}, f} \right), \nonumber \\
    H_{\text{II}} &= \lambda (t) H_{35} + (1-\lambda (t)) H_{34}  , \; \; \; t \in \left( t_{\text{II},i}, t_{\text{II}, f}  \right), \nonumber \\
    H_{\text{III}} &= \lambda (t) H_{23} + (1-\lambda (t)) H_{35} , \; \; \;  t \in \left( t_{\text{III},i}, t_{\text{III}, f}  \right). 
\end{align}

with $H_{ij} = -t_{ij} \chi^\dagger_{i, \uparrow} \chi_{j, \uparrow} + \text{h.c.}$, and in particular

\begin{align}
    H_{23} &= - 2 \abs{t_{23}} \cos \left( 2 \pi Q_1 + \varphi_{23} \right), \nonumber \\
    H_{34} &= - 2 \abs{t_{34}} \cos \left( 2 \pi J_2 + \varphi_{34} \right), \nonumber \\
    H_{35} &= - 2 \abs{t_{35}} \cos \left( 2 \pi Q_2 + 2\pi J_2 + \varphi_{35} \right).
\end{align}

In the above, we assume that at all stages $k = \text{I, II, III}$, $\lambda (t_{i,k}) = 0, \lambda (t_{f, k}) = 1$ and phases $\varphi_{ij}$ that depend on microscopic details. 

It is important to note that at all times the Hamiltonian realized commutes with $e^{i 2\pi Q_3}$, \emph{and} a second operator $\Sigma_{k = \text{I,II,III}}$ that depends on which stage of the braiding process we are in. Specifically, 
\begin{align}
    \Sigma_{\text{I}} &= e^{-i 2\pi J_3},
    \Sigma_{\text{II}} &= e^{i 2\pi J_1},
    \Sigma_{\text{III}} &= e^{i 2\pi Q_2} e^{- i 2\pi J_3},  
    \label{eq:symms}
\end{align}
such that 
\begin{align}
    e^{i2\pi Q_3} \Sigma_{k} = \Sigma_{k} e^{i2\pi Q_3} e^{i \frac{2\pi}{m'-n}}, \; \; \; \text{for} \; \; k = \text{I, II, III}. 
    \label{eq:degeneracybraid}
\end{align}

Eq.~(\ref{eq:degeneracybraid}) guarantees that any time, the ground state manifold remains $m'-n$-fold degenerate. Thus, the ground state degeneracy of the original $N = 2$ segment system remains in place throughout the braiding process. (Note that the prototype in Fig.~\ref{fig:prototype} shows this a $N = 3$ line defect which includes the $2$ additional zero modes nucleated from the vacuum.) 

It is also worth noting at this stage that a term such as $\chi_{i,B}^\dagger \chi_{j,B}$ does not change the form of the terms $H_{23}, H_{34}$ but introduces another term $H_{35}' = - 2 \abs{t'_{35}} \cos \left(2\pi Q_2 - 2\pi J_2 + \varphi'_{35} \right)$ that in general ruins the symmetry $\Sigma_{\text{III}}$; thus, for the braiding scheme to be topologically robust, it is necessary to ensure that such a term is exponentially weaker than $H_{35}$. This is naturally realized in our system as modes $A, \bar{A}$ are always brought closer in the braiding process than modes $B, \bar{B}$.

\subsubsection{Unitary transformation realized}

We now compute the unitary transformation realized as a consequence of the above machinations. To do so, we first note that at any instant, $e^{i 2\pi Q_3}$ commutes with the Hamiltonian and thus the charge $q_3$ (defined modulo $m'-n$) is conserved and can be used to label the instantaneous ground states. Thus, at each stage of the process, unitary $U_{\text{k}}$ describing the time evolution at stage $k = \text{I, II, III}$ transforms the ground state manifold from a set of degenerate initial states (marked by a subscript $i$) into a new set of final states (marked by a subscript $k$), with a phase that depends on $q_3$. Thus, 

\begin{align}
    U_{k} \ket{\psi^k_i (q_3)} = e^{i \gamma_k (q_3)} \ket{\psi^k_f (q_3)}, 
\end{align}

where $\ket{\psi^k_i (q_3)}, \ket{\psi^k_f (q_3)}$ are eigenstates of the Hamiltonian $H_k$ at the initial and final times $t_{k,i}, t_{k,f}$, respectively. It is useful to work with the choice $\ket{\psi^{\text{I}}_f (q_3)} = \ket{\psi^{\text{II}}_i (q_3)}$---the total phase accumulated in the entire cycle is then given by $\Gamma (q_3) \equiv \gamma_{\text{I}} (q_3) + \gamma_{\text{II}} (q_3) + \gamma_{\text{III}} (q_3)$. 

The phases $\gamma_k (q_3)$ of course depend on an arbitrary choice of phase for the eigenstates of the Hamiltonian at initial and final times. However, the total phase accumulated after a complete cycle does not depend on these details and is determined by the braiding statistics. To determine this phase, we additionally define another set of phases $\delta^k_{i} (q_3), \delta^k_f (q_3)$ satisfying 

\begin{align}
    \Sigma_k \ket{\psi_{k, i(f)} (q_3)} = e^{i \delta^k_{i(f)} (q_3)} \ket{\psi_{k, i(f)} (q_3+1)}
    \label{eq:rel1}
\end{align}

where we note that the symmetry operator $\Sigma_k$ sends $q_3 \rightarrow q_3 + 1$ as a consequence of its commutation relation with $e^{i 2\pi Q_3}$. Next, noting that $\Sigma_k$ commutes with $U_k$, we obtain the relation

\begin{align}
    U_k \Sigma_k \ket{\psi_{i} (q_3)} = e^{i \gamma_k (q_3)} \Sigma_k \ket{\psi_{f} (q_3)}
     \label{eq:rel2}
\end{align}

Eqs.~(\ref{eq:rel1},\ref{eq:rel2}) imply the relation between phases $\gamma_k (q_3), \gamma_k (q_3 + 1)$ 

\begin{align}
    \gamma_k (q_3 + 1) = \gamma_k (q_3) + \delta^k_f (q_3) - \delta^k_i (q_3)
    \label{eq:rel3}
\end{align}

Thus, clearly $\gamma_k (q_3)$ are related, and the total phase accrued in a cycle satisfies

\begin{align}
    \Gamma (q_3 + 1) -\Gamma (q_3) &=  \sum_{k = \text{I,II}} \left( \delta^{\text{k}}_f (q_3) - \delta^{\text{k+1}}_i (q_3) \right) \nonumber \\
    &+ \left( \delta^{\text{III}}_f (q_3) - \delta^{\text{I}}_i (q_3) \right),
    \label{eq:rel4}
\end{align}

where we have arranged the phase difference between ground states labaled by $q_3, q_3 + 1$ in terms of phases differences that are independent of the gauge choice of phases for the eigenstates of the Hamiltonians $H_{23}, H_{34}$ and $H_{35}$. Thus, to evaluate the topological phase accrued by each state, $\Gamma (q_3)$, it remains to consistently define eigenstates of the three Hamiltonians realized at the end/beginning of the three stages, and to compute the phases $\delta^k_{i (f)}$ defined via the action of the two symmetry operators $\Sigma_k$ relevant to that stage. We relegate these details to Appendix~\ref{app:braidingU} and quote the final result here---

\begin{align}
\Gamma (q_3) &= e^{i \frac{\pi}{m'-n} \left( q_3 - k/2 \right)^2}
\label{eq:q3}
\end{align}

for some odd integer $k$ that will generically depend on microscopic details. Finally, note that at the beginning (and end) of the cycle, the eigenstates have well defined charges $q_1, q_2, q_3$. Additionally, without loss of generality, we may assume i) we work in the ground state subspace defined by the total charge $q_1 + q_2 + q_3 \equiv 0$ (modulo $m'-n$), and ii) the vacuum state furnishes repeatedly a fixed (but undetermined) charge $q_1 = q_v$. Using these, we can replace $q_3$ in favor of $q_2$, the charge trapped between the interfaces $4,5$ that we are braiding. In total, we note that braiding modes $4,5$ via the process outlined yields a unitary 

\begin{align}
    \mathcal{U}_{45} &= \exp \left\{ - i \pi (m'-n) \left( \hat{Q}_2 - \frac{k_{4}}{2(m'-n)} \right) \right\}
\end{align}, 

for some odd integer $k_{4}$. Generalizing this result to arbitrary pairs of nearest neighbor zero modes, we obtain

\begin{align}
    \mathcal{U}_{2j, 2j+1} &= \exp \left\{ - i \pi (m'-n) \left( \hat{Q}_j - \frac{k_{2j}}{2(m'-n)} \right) \right\}, \nonumber \\
    \mathcal{U}_{2j-1, 2j} &= \exp \left\{ - i \pi (m'-n) \left( \hat{J}_j - \frac{k_{2j-1}}{2(m'-n)} \right) \right\}, \nonumber
    \label{eq:braidingU}
\end{align}

for odd integer $k_j$ that depend on microscopic details. 

\subsubsection{Yang-Baxter equations}

We can show that the above equations satisfy the Yang-Baxter equations which confirm that they form a representation of the braid group. In particular, it is clear from the commutation $\left[e^{i 2\pi \hat{Q}_i}, e^{i 2\pi \hat{J}_j} \right] = 0$ for $\abs{i-j} > 1$, and $\left[ e^{i 2\pi \hat{Q}_i}, e^{i 2\pi \hat{Q}_j}\right] = \left[ e^{i 2\pi \hat{J}_i}, e^{i 2\pi \hat{J}_j}\right] = 0$, that braiding pairs of zero modes that have no mode in common can be performed in either order. Thus,

\begin{align}
    \left[ \mathcal{U}_{j,j+1}, \mathcal{U}_{i,i+1} \right] = 0, \; \; \; \text{for} \; \; \; \abs{i-j} > 1
\end{align}

The non-trivial relation of the braid group 

\begin{align}
    \mathcal{U}_{j,j+1} \mathcal{U}_{j+1,j+2} \mathcal{U}_{j,j+1} = \mathcal{U}_{j+1,j+2} \mathcal{U}_{j,j+1} \mathcal{U}_{j+1,j+2}
    \label{eq:YB}
\end{align}

is also satisfied. To show this, we use the discrete Fourier expansion of the braiding unitaries. Using Eq.~(2.8) of Ref.~\cite{berndt1981determination}, we find for instance

\begin{align}
    \mathcal{U}_{23} &= \frac{1}{\sqrt{\abs{m'-n}}} \sum_{j = 0}^{\abs{m'-n}-1} e^{\frac{i\pi}{\abs{m'-n}} \left(j-1/2 \right)^2 - \frac{i \pi}{4}}e^{i 2\pi j Q_1}, \nonumber \\
    \mathcal{U}_{34} &= \frac{1}{\sqrt{\abs{m'-n}}} \sum_{j = 0}^{\abs{m'-n}-1} e^{\frac{i\pi}{\abs{m'-n}} \left(j-1/2 \right)^2 - \frac{i \pi}{4}}e^{i 2\pi j J_2}. \nonumber \\
    \label{eq:U234}
\end{align}

Here we have absorbed the undetermined even integers $k_{2}-1, k_{3}-1$ [associated with the braiding matrix as seen in Eq.~(\ref{eq:braidingU})] into the definition of $Q_1, J_2$ for convenience. Plugging this expansion into the left hand and right sides of Eq.~(\ref{eq:YB}), performing a change in variables, allows us to verify the Yang-Baxter relation involving the zero modes at interfaces $2,3$ and $4$. We reserve the details to Appendix~\ref{sec:YBproof}. This completes our discussion of the topological degeneracy in this system, the presence of non-Abelian zero modes, and their braiding properties. 

%In particular, we imagine nucleating a pair of non-Abelian anyon zero modes from the vacuum by nucleating a segment that pins the field $\theta_\sigma$ in a segment that otherwise pins $\phi_\sigma$. In the prototype shown in Fig.~\ref{fig:prototype} for instance, a vacuum resource can be furnished by pinching the zero modes realized at interfaces $5,6$ close to one another (thus effectively shrinking the line defect between interfaces $5,6$ to length zero) and unpinching/distancing them when required. This has the effect that the line defect between interfaces $1$ and $4$  

%\section{Material considerations and other details}

%We now briefly provide a discussion of the multi-valley systems that could serve as candidates to realizing the topological qubit proposed in this work, and the superlattice structures supporting charge density waves that may be appropriate. 

\section{Conclusions}
\label{sec:conclusions}

Non-Abelian anyonic zero modes have immense potential for serving as the basis for a topologically protected quantum computer and several proposals have been put forth to realize these elusive excitations in various materials and heterostructures. In this work, we introduce a new platform, that of quantum Hall valley ferromagnets, as a possible setting to realize such zero modes. Two possible advantages of this platform is that i) the physics discussed could be realized on silicon and other semiconducting materials that host multiple degenerate valleys and which have been developed on an industrial scale, and ii) the zero modes realized can be manipulated and braided using only local strain activated by piezoelectrics, thus eschewing the need for dynamical control of electromagnetic fields whose long range character can often lead to other complications such as stray fields.

This work takes inspiration from a whole host of proposals discussing how non-Abelian zero modes may be realized in defects in Abelian quantum Hall states. Specifically, our proposal is closest in spirit to the proposal of Ref.~\cite{barkeshli2014synthetic} of genon qubits, which involves a bilayer quantum Hall device hosting fractional Abelian quantum Hall states. Local charge density is depleted by gating, and interlayer tunneling is cleverly used to suture up the layers in a way as to produce an effectively higher genus surface for the electrons living in the bulk. This naturally leads to a degeneracy which grows exponentially in the genus of the surface, with the base being an integer that depends on the properties of the fractional quantum Hall state realized. 

In this proposal, valleys of a multivalley system take up the role of the layers, and line defects arise naturally between regions where different valleys are occupied (by virtue of a valley-ferromagnetic instability). The position of these line defects can be controlled by strain, which can gently bias the ferromagnetic instability towards favoring the occupation of one set of valleys over the other, and can thus be used to braid the realized zero modes in the scheme we discuss. The major distinction between the genon qubit proposal and the present proposal is due to the fact that the gapping perturbations (that `suture' the quantum Hall states across the line defect) arise as a consequence of interactions and not single particle tunneling. As we show, the line defects can be understood in terms of a two component Luttinger liquid with a gapped, electrically charged mode, and a gapless electrically neutral mode composed of linear combinations of electrons in all four valleys. It is the charge neutral mode that is then gapped using appropriate alternating perturbations, provided by a superlattice, that realizes the relevant topologically protected ground state subspace that we use for storing and manipulating quantum information. 

Several aspects of our proposal have been investigated experimentally. Various fractional quantum Hall states have been realized in silicon but require further exploration~\cite{dunford1995fqhe,lai2004two,lu2012fractional}, valley polarized fractional quantum Hall states were demonstrated on Bi(111) in Ref.~\cite{feldman2016observation} and line defects and their properties have been investigated~\cite{randeria2019interacting}, superlattices have been imposed on multivalley quantum Hall states~\cite{hossain2021bloch} (in AlAs to study the anisotropic mass of composite fermions at half filling). It would be a useful first step to numerically investigate, whether the kinds of fractional quantum Hall states theorized in this work to exhibit non-Abelian zero modes are actually realized in a realistic setup, and develop a more microscopic theory of line defects in this system. More generally, multivalley systems are a remarkable test bed for new strongly correlated electronic phases in the quantum Hall regime and we hope the present work provides further impetus to investigate the rich physics of these systems. 

%More generally, valley polarized fractional quantum Hall states were clearly demonstrated on Bi(111) in Ref.~\cite{feldman2016observation} and line defects and their properties have been investigated more recently in the case of integer filling of the bulk~\cite{randeria2019interacting} (where interactions are nevertheless required to enable valley symmetry breaking, and were also shown to be crucial in determining the spectral properties of the line defects). Superlattices have been imposed on multivalley quantum Hall states in AlAs to study the anisotropic mass of composite fermions at half filling. These systems are a remarkable test bed for new strongly correlated electronic phases in the quantum Hall regime and we hope the present work provides further impetus to investigate the rich physics of these systems. 

\section{Acknowledgements}
The author acknowledges fruitful discussions with Jason Alicea, Ajit Balram, Ravin Bhatt, Maissam Barkeshli, Michael Gullans, Netanel Lindner, Alexei Kitaev, Sid Parameswaran and Shivaji Sondhi. The author also acknowledges funding support from NSERC Grants RGPIN-2019-06465, and DGECR-2019-00011, an INTRIQ grant, and a Tomlinson Science Award from McGill University, and the hospitality of the Aspen Center for Physics, which is supported by National Science Foundation grant PHY-1607611, where some of the initial ideas in this work were presented and worked upon. 

\appendix

\section{Calculation of the braiding unitary}
\label{app:braidingU}

To compute the braiding unitary, we need to compute the phases $\gamma_k (q_3)$ accrued in the various stages $k = \text{I, II, III}$ of the time evolution. As outlined in the main text, in order to do so, we have to define eigenstates for the Hamiltonians $H_{23}, H_{34}, H_{35}$, along with some convention for the phase of these eigenstates. To obtain the path-invariant topologically robust phase accrued in the cycle, it is useful define phases $\delta^{k}_{i(f)} (q_3)$ which are defined through the relation Eq.~(\ref{eq:rel1}). Then, one can compute the total phase accrued $\Gamma_k (q_3) = \gamma_{\text{I}} (q_3) + \gamma_{\text{II}} (q_3) + \gamma_{\text{III}} (q_3)$ using Eqs.~(\ref{eq:rel3},\ref{eq:rel4}). Here we provide a computation of these phases by explicitly solving for the Hamitonians $H_{23}, H_{34}, H_{35}$, writing down their eigenstates, and determining the phases $\delta^{k}_{i(f)}$. 

As per the convention used in the rest of the test, we will work in a basis $\ket{q_2, q_3}$, and we assume that $q_1 = - q_2 - q_3 + x$, where $x$ is some fixed integer denoting the conserved total charge. In all cases this integer will be subsumed by the arbitrary tunneling phase and will thus not be explicitly stated. 

For $H_{23} = - 2 \abs{t_{23}} \cos \left( 2 \pi Q_1 + \varphi_{23} \right)$, the Hamiltonian is already diagonal in the assumed basis. The Hamiltonian and eigenstates read

\begin{align}
    &H_{23} = - 2 \abs{t_{12}} \cos \left( \frac{2\pi}{m'-n} \left( q_2 + q_3 \right) - \varphi_{23} \right) \nonumber \\
    &\ket{\psi^{\text{III}}_f (q_3)} = \ket{\psi^{\text{I}}_i (q_3)} = \ket{q_2 = -q_3 + k_{\text{I}}, q_3} 
\end{align}

Here $k_{\text{I}}$ is the integer nearest to $(m'-n) \varphi_{23} / 2\pi$. Note that the Hamiltonian may have $2(m'-n)$ degeneracies at special choices for $\varphi_{23}$; we assume that such choice of phase $\varphi_{23}$ is avoided. Then, it is sensible to label the lowest energy $m'-n$ eigenstates with the integer $q_3$. Here we denote both the eigenstates $\psi^{\text{III}}_{f} (q_3), \psi^{\text{I}}_i (q_3)$ at the end of the third stage, where the Hamiltonian $H_{23}$ is relevant, by the same state. 

For $H_{34} = - 2\abs{t_{34}} \cos \left( 2 \pi J_2 + \varphi_{34} \right)$, noting that the operator $e^{ \pm i 2 \pi J_2}$ sends $q_2 \mp 1$, we can write down the Hamiltonian in our chosen basis, and find eigenstates as follows---

\begin{align}
    &H_{23} = - \abs{t_{23}} \sum_{q2 = 0}^{m'-n'-1} e^{i \varphi_{34}} \ket{q_2} \bra{q_2 + 1} + \text{h.c.} \nonumber \\
    &\ket{\psi^{\text{I}}_f (q_3)} = \ket{\psi^{\text{II}}_i (q_3)} \nonumber \\
    &= \frac{1}{\sqrt{m'-n}}\sum_{j=0}^{m'-n-1} e^{i \frac{2\pi}{m'-n} j k_{\text{II}}} \ket{q_2 = - q_3 + j, q_3} 
\end{align}

where $k_{\text{II}}$ is the integer closest to $-(m'-n) \varphi_{34}/2\pi$. Here we note that the eigenstates are simply constant momentum states on the finite lattice of states $\ket{q_2}$. 

Finally, for $H_{35} = - 2 \abs{t_{35}} \cos \left( 2\pi Q_2 + 2\pi J_2 + \varphi_{35} \right)$, we note that the hopping for $\ket{q_2}$ to $\ket{q_2 \pm 1}$ is accompanied by a $q_2$-dependent phase $e^{ \mp i\frac{2\pi}{m'-n}q_2}$. In order to diagonalize this Hamiltonian, it is useful to define another set of states that absorb this phase as follows---

\begin{align}
    \ket{\overline{q_2}} = \ket{q_2} e^{-i \frac{\pi}{m'-n} q^2_2 - i \frac{\pi}{m'-n} q_2 }. 
\end{align}

The Hamiltonian in terms of these states reads 

\begin{align}
    H_{35} &= - \abs{t_{35}} \sum_{q_2 = 0}^{m'-n-1} e^{i \varphi_{35} + i \frac{2\pi}{m'-n} q_2} \ket{q_2 - 1} \bra{q_2} + \text{h.c.} \nonumber \\
    &= - \abs{t_{35}} \sum_{q_2 = 0}^{m'-n-1} e^{i \varphi_{35}} \ket{\overline{q_2 - 1}} \bra{\overline{q_2}} + \text{h.c.},
\end{align}

and the eigenstates can be read out as momentum eigenstates in this new basis. Reverting to the usual basis, we find

\begin{align}
    & \ket{\psi^{\text{II}}_f (q_3)} = \ket{\psi^{\text{III}}_i (q_3)} = \nonumber \\
    & \frac{1}{\sqrt{m'-n}}  \sum_{q_2 = 0}^{m'-n -1} e^{i \frac{2\pi}{m'-n} q_2 k_{\text{III}} - i \frac{\pi}{m'-n} (q^2_2 + q_2) } \ket{q_2,q_3} 
\end{align}

where $k_{\text{III}}$ is the nearest integer to $-(m'-n)\varphi_{35}/2\pi$. 

With the eigenstates in hand, it remains to compute the phases $\delta^k_{i,f} (q_3)$ defined through Eq.~(\ref{eq:rel1}). First, we note that the relevant symmetry operators at stages $k = \text{I, II, III}$ that increase $q_2 \rightarrow q_2 + 1$ are given as per Eq.~(\ref{eq:symms}) in the main text. Applying these operators to the relevant wave functions, we find

\begin{align}
    \delta^{\text{I}}_i (q_3) &= \delta^{\text{I}}_f (q_3) = \delta^{\text{II}}_f (q_3) = 0 \nonumber \\
    \delta^{\text{II}}_i (q_3) &= -\frac{2\pi}{m'-n} k_{\text{II}}, \nonumber \\
    \delta^{\text{III}}_i (q_3) &= \frac{2\pi}{m'-n} (k_{\text{III}} - 1), \nonumber \\
    \delta^{\text{III}}_f (q_3) &= \frac{2\pi}{m'-n} (k_{\text{I}} - q_3 - 1), 
\end{align}

from which the phases accrued during each state of the evolution can be found using Eq.~(\ref{eq:rel3}) to be 

\begin{align}
    \gamma_{\text{I}} (q_3) &= 0 \nonumber \\
    \gamma_{\text{II}} (q_3) &= \frac{2\pi}{m'-n} k_{\text{II}} q_3 \nonumber \\
    \gamma_{\text{III}} (q_3) &= - \frac{\pi}{m'-n} \left(q_3 - k_{\text{I}} + k_{\text{III}} - \frac{1}{2} \right)^2
\end{align}

where we have made certain gauge choices for $\gamma_{\text{I,II,III}} (q_3 = 0) $. The total phase accrued can now be computed by adding the phases in each of these states and gives rise to the result 

\begin{align}
    \Gamma (q_3) = \frac{2\pi}{m'-n} \left( q_3 - k_{\text{I}} + k_{\text{III}} -1/2 - k_{\text{II}} \right)^2. 
\end{align}

where we have ignored a $q_3$-independent piece. This leads to the final result quoted in Eq.~(\ref{eq:q3}). 

\section{Proof of satisfaction of Yang-Baxter equation}
\label{sec:YBproof}

For the proof, as noted in the main text, we restrict our attention to proving the Yang-Baxter equation is satisfied by apropriate braids on intefaces $2,3,4$. In particular, using equations Eq.~(\ref{eq:U234}) for the braiding unitaries $\mathcal{U}_{23}, \mathcal{U}_{34}$, we would like to prove

\begin{align}
    \mathcal{U}_{23} \mathcal{U}_{34} \mathcal{U}_{23} = \mathcal{U}_{34} \mathcal{U}_{23} \mathcal{U}_{34}. 
    \label{eq:YBeq}
\end{align}

For convenience, we will drop the subscript in the charges $Q_1, J_2$ in what follows. We can now expand the braiding unitaries on the left hand side of Eq.~(\ref{eq:YBeq}) in Fourier space to find 

\begin{widetext}
\begin{align}
    \text{L.H.S.} &= \sum_{k_1,k_2,k_3 = 0}^{m'-n-1} \frac{e^{-i 3\pi /4}}{(m'-n)^{3/2}} e^{\frac{i \pi}{m'-n} \left( (k_1 - 1/2)^2 + (k_2 - 1/2)^2 + (k_3 - 1/2)^2  \right) } e^{i 2\pi k_1 Q} e^{i 2\pi k_2 J} e^{i 2\pi k_3 Q} \nonumber \\
    &= \sum_{k_1,k_2,k_3 = 0}^{m'-n-1} \frac{e^{-i 3\pi /4}}{(m'-n)^{3/2}} e^{\frac{i \pi}{m'-n} \left( (k_1 + k_2 - 1/2)^2 + (k_3 - 1/2)^2 + 1/4 \right) } e^{i 2\pi k_2 J} e^{i 2\pi (k_1 + k_3) Q} \nonumber \\
    & \text{Now using the transformation $k_1 \rightarrow k_3 - k_1, k_2 \rightarrow k_2 + k_1, k_3 \rightarrow k_1$, we find} \nonumber \\
    \text{L.H.S} &= \sum_{k_1,k_2,k_3 = 0}^{m'-n-1} \frac{e^{-i 3\pi /4}}{(m'-n)^{3/2}} e^{\frac{i \pi}{m'-n} \left( (k_2 + k_3 - 1/2)^2 + (k_1 - 1/2)^2 + 1/4 \right) } e^{i 2\pi (k_1 + k_2) J} e^{i 2\pi  k_3 Q} \nonumber \\
    &= \sum_{k_1,k_2,k_3 = 0}^{m'-n-1} \frac{e^{-i 3\pi /4}}{(m'-n)^{3/2}} e^{\frac{i \pi}{m'-n} \left( (k_2 - 1/2)^2 + (k_1 - 1/2)^2 + (k_3 - 1/2)^2 \right) } e^{i 2\pi (k_1 J} e^{i 2\pi  k_3 Q} e^{i 2\pi k_2 J } \nonumber \\
    & = \text{R.H.S}
\end{align}
\end{widetext}

In the above, made use of the commutation relation $e^{i 2\pi a Q} e^{i 2\pi b J} = e^{i \frac{2\pi a b}{m'-n}} e^{i 2\pi b J} e^{i 2\pi a Q}$ for integer $a,b$ twice. This completes the proof showing that the briading unitaries satisy the Yang-Baxter equation.

\bibliographystyle{apsrev4-1}
\bibliography{nonabelian.bib}

\end{document}